\documentclass[fleqn,usenatbib]{mnras}

\usepackage{newtxtext,newtxmath}

\usepackage[T1]{fontenc}

\DeclareRobustCommand{\VAN}[3]{#2}
\let\VANthebibliography\thebibliography
\def\thebibliography{\DeclareRobustCommand{\VAN}[3]{##3}\VANthebibliography}

\usepackage{graphicx}	
\usepackage{amsmath}	
\usepackage{xcolor}
\usepackage{xspace}
\newcommand{\python}{\textsc{python}\xspace}
\newcommand{\afterglowpy}{\textsc{afterglowpy}\xspace}
\newcommand{\dynesty}{\textsc{dynesty}\xspace}
\newcommand{\bilby}{\textsc{bilby}\xspace}

\usepackage{gensymb}

\title[$H_0$ estimation of GW170817]{Potential biases and prospects for the Hubble constant estimation via electromagnetic and gravitational-wave joint analyses}

\author[G. Gianfagna et al.]{%
Giulia Gianfagna,$^{1,2}$\thanks{E-mail: giulia.gianfagna@inaf.it}
Luigi Piro,$^{1}$
Francesco Pannarale,$^{2,3}$
Hendrik Van Eerten,$^{4}$
Fulvio Ricci,$^{2,3}$
\newauthor
Geoffrey Ryan$^{5}$\\
$^{1}$INAF - Istituto di Astrofisica e Planetologia Spaziali, via Fosso del Cavaliere 100, 00133 Rome, Italy\\
$^{2}$Dipartimento di Fisica, Università di Roma "Sapienza", Piazzale A. Moro 5, I-00185, Roma, Italy\\
$^{3}$INFN Sezione di Roma, Piazzale A. Moro 5, I-00185, Roma, Italy\\
$^{4}$Department of Physics, University of Bath, Claverton Down, Bath, BA2 7AY, UK\\
$^{5}$Perimeter Institute for Theoretical Physics, 31 Caroline St. N., Waterloo, ON, N2L 2Y5, Canada
}

\date{Accepted 2024 January 15. Received 2023 December 20; in original form 2023 September 28}

\pubyear{2024}

\begin{document}
\label{firstpage}
\pagerange{\pageref{firstpage}--\pageref{lastpage}}
\maketitle

\begin{abstract}
GW170817 is a binary neutron star merger that exhibited a gravitational wave (GW) and a gamma-ray burst, followed by an afterglow.
In this work, we estimate the Hubble constant ($H_0$) using broad-band afterglow emission and relativistic jet motion from the Very Long Baseline Interferometry and Hubble Space Telescope images of GW170817. 
Compared to previous attempts, we combine these messengers with GW in a simultaneous Bayesian fit. We probe the $H_0$ measurement robustness depending on the data set used, the assumed jet model, the possible presence of a late time flux excess.  
Using the sole GW leads to a $20\%$ error ($77^{+21}_{-10}$\,$\rm km/s/Mpc$, medians, 16th-84th percentiles), because of the degeneracy between viewing angle ($\theta_v$) and luminosity distance ($d_L$). 
The latter is reduced by the inclusion in the fit of the afterglow light curve, leading to $H_0=96^{+13}_{-10}$\,$\rm km/s/Mpc$, a large value, caused by the fit preference for high viewing angles due to the possible presence of a late-time excess in the afterglow flux. Accounting for the latter by including a constant flux component at late times brings $H_0=78.5^{+7.9}_{-6.4}$ $\rm km/s/Mpc$. Adding the centroid motion in the analysis efficiently breaks the $d_L-\theta_v$ degeneracy and overcome the late-time deviations, giving  $H_0 = 69.0^{+4.4}_{-4.3}$\,$\rm km/s/Mpc$ (in agreement with \textit{Planck} and SH0ES measurements) and $\theta_v = 18.2^{+1.2}_{-1.5}$ deg. This is valid regardless of the jet structure assumption. Our simulations show that for next GW runs radio observations are expected to provide at most few other similar events. 


\end{abstract}




\section{Introduction}
\label{sec:intro}

The $\Lambda$CDM model is the currently adopted standard model of cosmology. Great effort has been put into the estimation of one of its key parameters, the Hubble constant $H_0$, the current expansion rate of the Universe. 
The $\Lambda$CDM model calibrated with data from the \textit{Planck} mission, that is, from early-Universe physics, predicts the Hubble constant to 1\% precision: $67.4\pm0.5 \ \rm km\ s^{-1}Mpc^{-1}$ \citep{Planck2020} (we quote medians and 68\% credible intervals).
However, $H_0$ can also be empirically measured locally ($z<1$), in the late-time Universe. The latter kind of measurements, such as from SH0ES (Supernovae $H_0$ for the Equation of State, \citealt{Riess2019}) and H0liCOW ($H_0$ lenses in COSMOGRAIL's Wellspring, \citealt{Wong2019}), favour larger values of $H_0$: $74.0\pm1.4 \ \rm  km\ s^{-1}Mpc^{-1}$ and $73.3\pm1.8 \ \rm  km\ s^{-1}Mpc^{-1}$, respectively. Thus, the early-Universe data seem to be consistently predicting a low value of $H_0$, while
the late-time Universe data a higher one, leading to a more than 3$\sigma$ discrepancy \citep[see for an extensive discussion][]{Verde2019}.

A way to solve this discrepancy is to measure the Hubble constant through an independent method, using, for example, gravitational waves (GWs), where the distance is directly estimated from fitting the waveform, relying only on the general theory of relativity. This estimation does not depend on cosmic distance ladders. GWs can determine the Hubble constant if the redshift is provided by an electromagnetic (EM) counterpart, for example by a kilonova \citep{Taylor2012, Feeney2021}. This is the so-called standard sirens method \citep{Holz2005, Nissanke2010}. Even when a unique counterpart cannot be identified, the redshifts of all the potential host candidates can be incorporated in the analysis, when the localization volume is sufficiently small. This is not as constraining as the first scenario, but it is still informative, once many detections are available. In this case, more than 50 binary neutron stars are needed to reach a 6\% $H_0$ measurement \citep{Chen2018}. The same holds also for GWs emitted by binary black holes, even if the localization volumes are usually much larger than for binary neutron stars. In this case $\sim500$ events are needed to reach a precision $<7\%$ on $H_0$ \citep{Chen2022, bom2023}. 

The main problem of the standard sirens method is the degeneracy between the luminosity distance and inclination (the angle between the total angular momentum and the line of sight) estimated from GWs. They are measured from the amplitude of the two GW polarizations. At small inclinations, the cross and plus polarizations have nearly the same amplitude, but the larger the inclination, the more they decrease and start to differ \citep{Usman2019}. This means that the GW signal is strongest at small inclinations (face-on or face-off), but, in these cases, we cannot measure distance and inclination separately.
Therefore, associated EM observations can lead to a tighter measurement of $H_0$ by providing additional constraints on the inclination.

The first measurement of $H_0$ using GWs was obtained with the first binary neutron star merger observation GW170817, by combining the distance from the GW signal and the recession velocity of the host galaxy, resulting in $H_0$ of $74^{+16}_{-8} \ \rm km\ s^{-1}Mpc^{-1}$ \citep{Abbott2017_dl}. GW170817 was detected by the two Advanced LIGO detectors \citep{AdvancedLIGO2015} and Advanced Virgo \citep{AdvancedVIRGO2015} on August 17, 2017 \citep{Abbott2017_1}. It was identified as the collision of two neutron stars, which is theoretically expected to be followed by a highly relativistic jet, from which a gamma-ray burst (GRB) of short duration
($\lesssim 2\,$s) is produced \citep{Blinnikov1984, Paczynski1986, Eichler1989, Paczynski:1991aq, Narayan1992, Rhoads1997, Piran2005, nakar2007, Berger2014, Nakar2020, Salafia2022}.
This was proven by the joint detection of the GW event and of the short, hard burst GRB 170817A \citep{Goldstein2017, Savchenko2017, Abbott2017_dis}; then observations in the X-ray \citep{Troja2017_Nature} and, later, radio frequencies \citep{Hallinan2017} showed the afterglow emission. These observations are consistent with a short GRB viewed off-axis \citep[e.g.,][]{Troja2017_Nature, Margutti2017, Haggard2017, Finstad2018, Alexander2018, gill2018, Dobie2018, Granot2018, DAvanzo2018, Lazzati2018, Lyman2018, Margutti2018, Mooley2018, Troja2018, Fong2019, Hajela2019, Lamb2019, Wu2019, Piro2018, Ryan2020, Troja2020,  troja2021, Takahashi2021, Hajela2022, Gianfagna2023, McDowell2023, Hayes2023}. Moreover, radio observations that measure the superluminal motion of the jet centroid in radio and optical images were performed \citep{Mooley2018, Ghirlanda2019, Mooley2022}.

The EM information on the inclination derived from the afterglow and the relativistic jet motion of GW170817 allow us to improve the Hubble constant measurement for the reason stated above \citep[see also][for a review]{Bulla2022}. 
The common practice is to use the GW analysis results (posterior) for inclination and luminosity distance, and apply these as \textit{a priori} information on the inclination obtained by fitting the EM data sets (or the other way around). This can be done using Bayesian analysis. 
The results retrieved in this way run from low values such as $H_0 = 66.2^{+4.4}_{-4.2} \ \rm km\ s^{-1}Mpc^{-1}$, from \citet{Dietrich2020}, who fit the kilonova emission and the jet centroid motion, and $H_0 = 69.5 \pm 4 \ \rm km\ s^{-1}Mpc^{-1}$, from \cite{Wang2021}, who use the afterglow emission, to high values such as ${H}_{0}=75.5^{+11.6}_{-9.6} \ \rm km\ s^{-1}Mpc^{-1}$, by \citet{Guidorzi2017}, who fit the afterglow up to 40 days from the merger. 
\citet{Wang2023} estimate ${H}_{0}=71.80^{+4.15}_{-4.07} \ \rm km\ s^{-1}Mpc^{-1}$, modelling the jet with hydrodynamic simulations, including also a sub-relativistic kilonova outflow.
\cite{Palmese2023} use the same model as \citet{Wang2023}, fitting the afterglow, but including \textit{a priori} information on the Lorentz factor from the jet centroid motion. They find ${H}_{0}=75.46^{+5.34}_{-5.39} \ \rm km\ s^{-1}Mpc^{-1}$. 
\citet{hotokezaka2018} fit the afterglow and the jet centroid motion, finding ${H}_{0}=68.9^{+4.7}_{-4.6} \ \rm km\ s^{-1}Mpc^{-1}$. In general, the smaller is the viewing angle, the higher is the luminosity distance (because of their degeneracy), the lower is $H_0$. 

In general, the best approach to estimate $H_0$ is to include all the available information about the analyzed event, so, in case of GW170817, the GW, the afterglow light curve, the jet centroid motion and the kilonova. However, even analyses including only some of these are useful, especially to quantify possible systematics, that will count as reference for future events. In the case of GW170817, the afterglow light curve and the centroid motion tend to give contrasting results, as will be shown in this work, mainly because of the late time data points in the light curve, which seem to be showing a flux excess. Therefore, the use of the complete available information is even more important.

At present, the aforementioned EM-informed $H_0$ measurements are a factor 2 more precise than the first standard-siren measurement for GW170817 that fitted GW data only \citep{Abbott2017_dl}. For this reason, this method is very compelling. However, there are potential systematics that should be addressed. 
An open issue, as outlined in \cite{Nakar2021}, is the sensitivity of EM-derived parameters, as the inclination, on the assumed jet structure. A related problem is the presence of deviations from the assumed model due to a possible flux excess at late times.
In this work, we assess these issues with a comprehensive approach. We estimate the Hubble constant exploiting the GW, the broad-band afterglow and the centroid motion of the relativistic jet of the GW170817 event. We test the sensitivity of the results on the jet structure, and check for potential biases, both due to the jet model assumption and to the possible presence of an excess at late times in the afterglow. We fit the GW, the afterglow and the centroid motion data sets simultaneously using a Bayesian approach \citep{Gianfagna2023} and compare the results obtained fitting only afterglow and GW data, and then including also the centroid motion of the relativistic jet. We focus on the degeneracy between the viewing angle and the luminosity distance. As already stated above, because of this degeneracy, the standard sirens method at present cannot give a Hubble constant estimation at the \textit{Planck} or SH0ES level of precision. Here we study how we can break this degeneracy including different types of EM messengers. We also test the robustness of the derived $H_0$ by implementing the different jet models on real data and simulations.

In Section\,\ref{sec:data} we present the data sets used in this work,  analysed following the method presented in Section\,\ref{sec:methods}. In Section\,\ref{sec:results} we show the results that we obtained, both for the energetics, microphysics and geometry of the event, and for the Hubble constant. Finally in Section\,\ref{sec:conclusions} we summarize our conclusions.

\section{Data}
\label{sec:data}

This work used three data sets pertaining to the GW170817 event and analyzes them simultaneously. These are the broad-band afterglow emission, the centroid position of the jet as a function of time, and the GW strain timeseries.

\begin{figure}
    \includegraphics[width=0.5\textwidth]{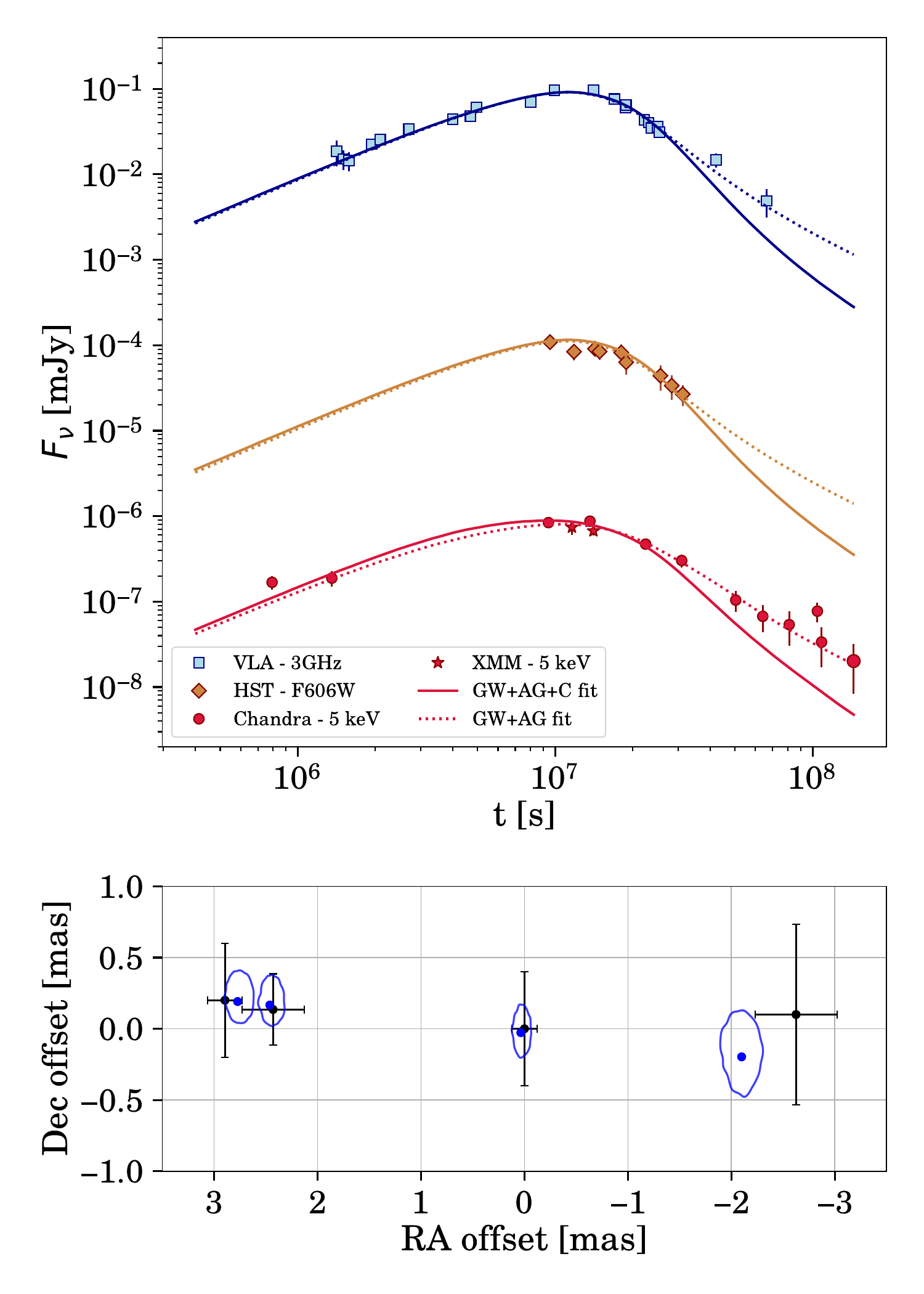}
    \caption{Top panel. Broad-band afterglow of GW170817: data and fits. From bottom to top, red points refer to the X-ray observations by \textit{Chandra} and \textit{XMM} at 5\,keV, orange ones to observations by \textit{HST}, F606W filter, in the optical band, and blue ones to observations in the radio band from VLA (Very Large Array) at 3\,GHz. The continuous and dotted lines represent the fit of the GW, broad-band afterglow, and centroid motion (GW+AG+C) data and of the GW and afterglow (GW+AG) data, respectively.  For sake of simplicity, the fit for the radio band is plotted only for the observations at 3\,GHz, but it is not limited to this single frequency. Bottom panel. Centroid motion of the relativistic jet from HST and VLBI images at 8 (negative RAs), 75, 206 and 230 days \citep{Mooley2018, Ghirlanda2019}. The blue dots represents the positions predicted by the model, the blue contours represent the 68\% probability region.}
     \label{Fig:lightcurve_gw170817}
\end{figure}

Regarding the afterglow emission, we include in the analysis data in the X-ray (\textit{Chandra} and \textit{XMM}), radio (frequencies from 0.7 to 15 GHz for VLA, ATCA, uGMRT, eMERLIN, MeerKAT) and optical (Hubble Space Telescope, HST) published in \citet{Troja2017_Nature, Fong2019, makhathini2021, troja2021, GCN_Chandra}, see also Fig.\,\ref{Fig:lightcurve_gw170817}.

We also include the centroid motion of the relativistic jet, visible in optical and radio images \citep{Mooley2018, Ghirlanda2019, Mooley2022}. For this analysis, we use the positions and uncertainties of the data points from VLBI (Very Long Baseline Interferometry) at 75, 206, 230 days reported in \citet{Mooley2018, Ghirlanda2019}, and from HST at 8 days \citep{Mooley2022}. For the latter we use the positions (right ascension, RA, and declination, Dec) and their statistical uncertainties, to which we add in quadrature the two systematic uncertainty contributions to take into account the different reference frame of the optical and radio images \citep[as in][]{Mooley2022}. 

The GW data of GW170817 are publicly available at the GW Open Science Center\footnote{\href{https://www.gw-openscience.org/}{www.gw-openscience.org}}\citep{LIGOScientific:2019lzm}. We use the cleaned version of the strain data, where the glitch discussed in \citet{Abbott2017_1} has been removed. 

\section{Joint analysis of electromagnetic and gravitational-wave data}
\label{sec:methods}

We use the Bayesian inference to process the GW and EM data. The two domains can be joined in one analysis as the models describing the two emissions (EM and GW) have parameters in common, namely the viewing angle and the luminosity distance. We perform three fits, one including only GWs, one folding GWs and the afterglow emission, and one including also the jet centroid motion data set. In this Section we describe the GW and the EM models, along with the joint fit method \citep[see also][]{Gianfagna2023}.

\subsection{Electromagnetic and gravitational-wave models}
\label{sec:models}

\subsubsection{Afterglow light curve and centroid motion}
\label{subsubsec:afterglow_model}

We model the broad-band afterglow light curve using the \python package \afterglowpy \citep{Ryan2020}. 
The observer frame flux of synchrotron radiation is estimated for various jet geometries. In this work we we use a Gaussian structured jet model, where the energy drops according to $E(\theta) = E_0 \exp(-\theta^2 / 2\theta_c^2)$, up to a truncating angle $\theta_w$. $E_0$, $\theta_c$ and $\theta_w$ are free parameters in the fit, representing the on-axis isotropic equivalent kinetic energy of the blast wave, the jet opening angle and the jet total angular width. We also use a power law jet model, where the energy is given by $E(\theta) = E_0 (1+(\theta^2/(b\theta_c^2))^{-b/2}$, where $b$ is the power law index.
The electrons are shock-accelerated and emit synchrotron radiation, with an energy distribution given by a power law with slope $-p$, the fraction of their post-shock internal energy is $\epsilon_e$, while the fraction of post-shock internal energy in the magnetic field is denoted by $\epsilon_B$. Furthermore, the circumburst medium number density $n_0$, the viewing angle $\theta_v$, between the jet axis and the line of sight, and the luminosity distance $d_L$ are also free parameters. The participation fraction $\chi_{N}$ is fixed to 1.0.

In order to model also the jet centroid motion, we use an extended version of \afterglowpy \citep{Ryan2023, Vaneerten2023}, where the afterglow image centroid position and sizes can be estimated. The imaging plane is perpendicular to the line of sight of the observer, and the centroid position and sizes are computed as an intensity-weighted quantity. The outputs of the model that we use in this work are the centroid position in the sky (RA and Dec) and the flux expected at each particular time. At 8 days, in the optical, only the kilonova emission is visible, and not the afterglow. For this reason, we place a $5\sigma$ upper limit for the optical flux of $4\times10^{-5}$ mJy. The parameters are the same as above, with an extra three parameters: $\rm RA_0, Dec_0$, which represent the jet origin in the sky image, and the position angle PA, which is the orientation of the jet direction in the image. 

The prior probability distributions are reported in Table \ref{table:priors}. The prior for the viewing angle $\theta_v$ is isotropic, meaning a sinusoidal distribution from 0$\degree$ to 90$\degree$ (uniform in cosine). For the luminosity distance, we use a uniform-in-volume prior ($\propto d_L^2$) from 1 to 75 Mpc, which distributes mergers uniformly throughout a Euclidean universe.

\begin{table}
    \caption{Prior probability distributions for the shared, the EM and GW fitted parameters.}
    \begin{tabular}{ccc} 
    \hline
    \noalign{\smallskip}
    Parameter & Prior functional form & Bounds \\
    \noalign{\smallskip}
    \hline
    \noalign{\smallskip}
        $d_{\rm L}$ [Mpc] & $\propto d_L^2$ & [1, 75] \\ 
        $\theta_{\rm v}$ [$\deg$] & sin($\theta_{\rm v}$) & [0, 90] \\
        \hline
        $\log_{10} E_0/\textrm{erg}$ & Uniform & [49, 56] \\
        $\theta_c$ [$\deg$] & Uniform & [0, 90]  \\
        $\theta_W$ [$\deg$] & Uniform & [0, 90] \\
        $\log_{10}n_0$ &  Uniform & [-7, 2] \\
        $p$  &  Uniform & [2, 3] \\
        $\log_{10} \epsilon_{\rm e}$ & Uniform & [-5, 0]  \\
        $\log_{10} \epsilon_{\rm B}$ & Uniform & [-5, 0]  \\
        RA$_0$ [mas] & Uniform & [-10, 10] \\
        Dec$_0$ [mas] & Uniform & [-10, 10] \\
        PA [$\deg$] & Uniform & [0, 360) \\
        \hline
        $ \mathcal{M} [M_\odot]$ & Uniform & [1.18, 1.21] \\
        $q$ & Uniform & [0.125, 1]  \\
        $a_1$ & Uniform & [0, 0.05] \\
        $a_2$ & Uniform & [0, 0.05] \\
        $\theta_1$ [$\deg$]& sin($\theta_1$) & [0, 180] \\
        $\theta_2$ [$\deg$]& sin($\theta_2$) & [0, 180] \\
        $\phi_{1,2}$ [$\deg$] &  Uniform & [0, 360] \\
        $\phi_{\rm JL}$ [$\deg$] & Uniform & [0, 360] \\
        $\psi$ [$\deg$] & Uniform & [0, 180]  \\
        $\Lambda_1$ & Uniform & [0, 5000] \\
        $\Lambda_2$ & Uniform & [0, 5000] \\
    
    \noalign{\smallskip}
    \hline
\end{tabular}
\label{table:priors}
\end{table}

\subsubsection{Gravitational waves}
\label{subsubsec:gw_model}

We use the {\ttfamily IMRPhenomPv2\_NRTidal} waveform approximant \citep{Hannam2014, Dietrich2017, Dietrich2019_2, Dietrich2019} to model GWs from binary neutron star mergers. The intrinsic parameters (the source physical parameters that shape the emitted signal) used by this model refer to the masses, the spins and the tidal deformabilities of the two neutron stars. The two component masses $m_1$ and $m_2$, for which we follow the common convention $m_1 \geq m_2$, will be quoted as the chirp mass \citep{Finn1993, Blanchet1995, Cutler1994, Poisson1995},
\begin{equation}
    \mathcal{M} = \frac{(m_1 m_2)^{3/5}}{(m_1 + m_2)^{1/5}}\,,
\end{equation}
and the mass ratio $q = m_2/m_1 \leq 1$.
The components of the dimensionless spin angular momenta of each neutron star, $\mathbf{a}_1$ and $\mathbf{a}_2$, constitute six additional parameters, which are: $a_1$ and $a_2$, the dimensionless spin magnitudes; $\theta_1$ and $\theta_2$, the tilt angles between the spins and the orbital angular momentum; $\phi_{1,2}$, the azimuthal angle separating the spin vectors; and $\phi_{JL}$, the opening angle of the cone of precession of the orbital angular momentum about the system's total angular momentum. The tidal deformability of each star is described by the dimensionless parameters $\Lambda_1$ and $\Lambda_2$.

The extrinsic parameters (that further shape the observed GW signal) in this model are the RA and Dec of the source (i.e., its sky position), the luminosity distance $d_L$ of the source, the inclination angle $\theta_{\rm JN}$ between the total angular momentum of the binary and the line of sight from the source to the observer, the polarization angle $\psi$, and the phase and time of coalescence. In this work, we fix the sky-position (RA and Dec) of the source to the one of AT 2017gfo \citep{Abbott2017_position}. The GW RA and Dec correspond to the $\rm RA_0, Dec_0$ parameters of the centroid motion model (Section\,\ref{subsubsec:afterglow_model}); however, the precision on RA and Dec from the GW data does not reach the mas level, as instead do $\rm RA_0$ and $\rm Dec_0$, so 
the analysis does not benefit from promoting the RA and Dec of the GW and the EM models to common, free parameters.
Moreover, we do not report the time of coalescence in the results, as this is of little interest in the context of our study, and we marginalize over the phase of coalescence. The latter marginalization is justified by the small spin magnitudes (see Table\,\ref{table:priors}), and hence the negligible precession effects \cite{Romero-shaw2020}. 

Given that the GRB jet develops around the total angular momentum, the inclination angle $\theta_{\rm JN}$ and the viewing angle $\theta_v$ introduced in Sec.\,\ref{subsubsec:afterglow_model} are essentially the same quantity, and thus a common parameter of the GW and EM domains. More precisely, in the case of GW170817, the two angles are supplementary \citep[see Eq.(1) in][]{Gianfagna2023}. The other parameter shared by the GW and EM domains is the luminosity distance. This implies that there are 23 parameters when addressing the GW and EM domains in our approach. 

The priors for the intrinsic and extrinsic GW parameters are set as in \citet{Romero-shaw2020}, in the case of the "Low Spin" analysis, see Table\,\ref{table:priors}.

\subsection{Joint fit}
\label{sec:methods-joint}

We use Bayesian inference to analyse jointly the data from the GW and EM domains, which we denote as $d_{GW}$ and $d_{EM}$, respectively \citep[the same methodology is presented in][]{Gianfagna2023}. 
The three main components are: a \textit{prior distribution}, which models the available knowledge about a given parameter before data collection in a statistical distribution; the \textit{likelihood function}, which encloses the information about the parameter from observed data; the \textit{posterior distribution}, which combines the prior distribution and the likelihood function using the Bayes theorem.
Thus, the multi-dimensional posterior probability distribution for our set of parameters $\vec{\vartheta}$ is:
\begin{equation}
   p(\vec{\vartheta} |d_{\textrm{EM}}, d_{\textrm{GW}} ) \equiv \frac{\mathcal{L}_{\textrm{EM+GW}}(d_{\textrm{EM}}, d_{\textrm{GW}}| \vec{\vartheta}) \pi(\vec{\vartheta})}{\mathcal{Z}_{\vec{\vartheta}}}
\label{Eq:bayes_th_joint}
\end{equation}
where $\mathcal{L}_{\textrm{EM+GW}}(d_{\textrm{EM}}, d_{\textrm{GW}}| \vec{\vartheta})$ is the likelihood function that folds the EM and GW domains,  $\pi(\vec{\vartheta})$ is the multi-dimensional prior probability distribution for our parameters, and $\mathcal{Z}_{\vec{\vartheta}}$ is the Bayesian evidence. This is obtained by marginalizing the joint likelihood over the GRB and GW parameters:
\begin{equation}
    \mathcal{Z}_{\vec{\vartheta}} = \int \mathcal{L}_{\textrm{EM+GW}}(d_{\textrm{EM}}, d_{\textrm{GW}}| \vec{\vartheta}) \pi(\vec{\vartheta}) d\vec{\vartheta}\,.
\end{equation}

When the two data sets are independent, as is the case here, the likelihood $\mathcal{L}_{\textrm{EM+GW}}$ \citep[see also][]{Fan2014, Biscoveanu2020} is simply given by the product of the EM and GW likelihoods
\begin{equation}
    \mathcal{L}_{\textrm{EM+GW}}(d_{\textrm{EM}}, d_{\textrm{GW}}| \vec{\vartheta}) = \mathcal{L}_{\textrm{EM}}(d_{\textrm{EM}}| \vec{\vartheta}) \times \mathcal{L}_{\textrm{GW}}(d_{\textrm{GW}}| \vec{\vartheta})\,.
\label{Eq:likelihood}
\end{equation}
The EM and GW likelihoods are both Normal distributions. The GW likelihood function is defined in, e.g., \citet{Finn1992, Romano2017, Romero-shaw2020}; in this likelihood, both the data and the model are expressed in the frequency domain. 

In the EM case, when only the afterglow is folded with the GW data (GW+AG), the likelihood function is proportional to $\exp (-{\chi }^{2}/2)$, where ${\chi }^{2}$ is given by the comparison between the expected flux and the entire
broadband set of afterglow data. 

In the case of the fit of the afterglow, centroid motion and GW strain (GW+AG+C), we assume the afterglow and the centroid motion data sets to be independent, being taken with different telescopes and at different times. The centroid data set includes the positions (RA and Dec) at each time and their respective fluxes, we take the data point at 8 days as the position of the merger. We assume the likelihood function to be a multivariate Normal distribution, where the expected centroid positions and fluxes from the model are compared with the three offset positions (RA and Dec) and the corresponding flux measurements. Moreover, we assume the covariance matrix to be diagonal (see \citealt{Ryan2023} for more details).
We place the centre of the centroid motion reference system at the positions corresponding to the observations at 75 days, as in \citet{Mooley2018, Ghirlanda2019, Mooley2022}. 

We use the Bayesian inference library \bilby \citep{bilby_paper, pbilby_paper} and the dynamic nested sampling package \dynesty \citep{Speagle2020} to simultaneously fit the EM and GW data sets. We use 2000 live points and multiple bounding ellipsoids as bounding strategy. The corner plots are created with the \textsc{corner} package \citep{corner}.

\subsection{Hubble constant estimation}
\label{subsubsec:H0_theory}

At small redshifts, as in the GW170817 case, the luminosity distance does not depend on the cosmological model, so the Hubble constant can estimated from
\begin{equation}
    v_H = H_0 \cdot d_L \,,
\label{eq:vH}
\end{equation}
where $v_H$ is the local "Hubble flow" velocity, in this case at the position of GW170817, and $d_L$ is the luminosity distance to the source. We follow the same procedure as \citet{Abbott2017_dl}, assuming a Normal distribution for $v_H = 3017\pm166 \ \rm km \ s^{-1}$.

\section{Results and discussion}
\label{sec:results}

We assume a Gaussian jet profile throughout the work, with the exception of Sec.\,\ref{subsec:PL_jet}, where we assume a power law profile, in order to test the sensitivity of the results on the jet model.

\renewcommand{\arraystretch}{1.5}
\begin{table*}
    \caption{Fit results for GW170817 for a Gaussian (GJ) and a power law jet (PLJ). We report the medians and the 16th-84th percentiles. In the second column we report the results for the GW-only fit, in the third and fifth columns the results of the fit of the broad-band afterglow and the GW, in the fourth and sixth columns the results of the joint fit of broad-band afterglow, centroid motion and GW. The two last columns provide the results of the GW+AG+C fit with a constant component of the type $F_{\nu, c} = 10^c$ modelling the late time emission.}
    \begin{tabular}{cccccccc} 
    \hline
    \noalign{\smallskip}
    Parameter & GW-only & GW+AG & GW+AG+C  & GW+AG & GW+AG+C & GW+AG & GW+AG+C\\
              &         & GJ & GJ & PLJ & PLJ & GJ + Constant & GJ + Constant\\    
    \noalign{\smallskip}
    \hline
    \noalign{\smallskip}
        $\log_{10} E_0$ & & $52.31^{+0.82}_{-0.80}$ & $54.50^{+0.28}_{-0.33}$ &  $52.12^{+0.78}_{-0.85}$ & $54.0^{+0.30}_{-0.32}$ & $52.81^{+0.90}_{-0.86}$ & $54.81^{+0.30}_{-0.35}$ \\
        $\theta_c$ [$\deg$] & & $7.73^{+0.86}_{-0.80}$ &  $2.85^{+0.24}_{-0.20}$ &  $5.57^{+0.69}_{-0.62}$  & $2.18^{+0.20}_{-0.16}$ & $5.37^{+0.97}_{-0.87}$ & $2.64^{+0.20}_{-0.18}$ \\
        $\theta_W$ [$\deg$] & & $57^{+19}_{-19}$ & $52^{+23}_{-23}$ &  $58^{+18}_{-18}$  &  $50^{+23}_{-25}$ & $52^{+22}_{-21}$ & $52^{+23}_{-23}$ \\
        $\log_{10}n_0$ & & $-0.68^{+0.80}_{-0.80}$ & $-1.93^{+0.34}_{-0.39}$ &  $-0.37^{+0.77}_{-0.84}$  &  $-2.40^{+0.40}_{-0.35}$ & $-1.39^{+0.89}_{-0.89}$ & $-1.86^{+0.35}_{-0.39}$ \\
        $p$  & & $2.11^{+0.01}_{-0.01}$ & $2.11^{+0.01}_{-0.01}$ &  $2.12^{+0.01}_{-0.01}$  &  $2.12^{+0.01}_{-0.01}$ & $2.12^{+0.01}_{-0.01}$ & $2.12^{+0.01}_{-0.01}$ \\
        $\log_{10} \epsilon_{\rm e}$ & & $-1.65^{+0.71}_{-0.73}$ & $-3.45^{+0.28}_{-0.24}$ &   $-1.34^{+0.74}_{-0.69}$ &  $-2.72^{+0.30}_{-0.25}$ & $-1.89^{+0.76}_{-0.79}$ & $-3.64^{+0.30}_{-0.24}$ \\
        $\log_{10} \epsilon_{\rm B}$ & & $-3.78^{+0.80}_{-0.80}$ & $-3.89^{+0.34}_{-0.29}$ &   $-3.83^{+0.82}_{-0.77}$ &  $-3.43^{+0.35}_{-0.28}$ & $-3.63^{+0.85}_{-0.88}$ & $-4.06^{+0.37}_{-0.32}$ \\
        $b$ &   &   &   & $7.5^{+1.6}_{-1.1}$ & $10.8^{+0.7}_{-1.0}$ &   & \\
        $c_{\rm radio}$ &   &  &   &  & & $-2.99^{+0.23}_{-0.20}$ & $-2.88^{+0.24}_{-0.26}$ \\    
        $c_{\rm optical}$ &  &   &   &  &  &$-5.25^{+0.23}_{-0.22}$ & $-5.24^{+0.24}_{-0.23}$ \\    
        $c_{\rm X-rays}$ &  &   &   &  &  & $-7.48^{+0.05}_{-0.03}$ & $-7.47^{+0.08}_{-0.10}$ \\    
        RA$_0$ [mas] & & & $-2.1^{+0.2}_{-0.2}$ &   &  $-2.0^{+0.2}_{-0.2}$ &  & $-2.2^{+0.2}_{-0.2}$ \\
        Dec$_0$ [mas] & & & $-0.2^{+0.3}_{-0.3}$ &  &  $-0.2^{+0.3}_{-0.3}$  & & $-0.2^{+0.3}_{-0.3}$ \\
        PA [$\deg$] & & & $85^{+4}_{-3}$ &  &  $85^{+4}_{-3}$  & & $85^{+5}_{-3}$ \\
        \hline
        $d_{\rm L}$ [Mpc] & $39.2^{+5.4}_{-8.6}$  & $31.3^{+3.0}_{-3.6}$ & $43.7^{+1.4}_{-1.4}$ &  $23.7^{+3.8}_{-3.4}$  & $43.0^{+1.4}_{-1.4}$  & $38.6^{+2.5}_{-3.0}$ & $44.3^{+1.4}_{-1.3}$ \\ 
        $\theta_{\rm v}$ [$\deg$] & & $50.1^{+5.1}_{-5.4}$ & $18.2^{+1.2}_{-1.5}$ &  $62.7^{+5.0}_{-4.3}$  &  $19.8^{+1.3}_{-1.8}$  & $35.2^{+5.7}_{-6.2}$ & $17.2^{+1.1}_{-1.2}$ \\
        $\theta_{\rm JN}$ [$\deg$]&  $146^{+16}_{-18}$ & $129.9^{+5.1}_{-5.4}$ & $161.8^{+1.2}_{-1.5}$ &  $117.3^{+5.0}_{-4.3}$ & $160.2^{+1.3}_{-1.8}$ & $144.8^{+5.7}_{-6.2}$ & $162.8^{+1.1}_{-1.2}$ \\
        \hline
        $ \mathcal{M} [M_\odot]$ & $1.1975^{+0.0001}_{-0.0001}$ & $1.1975^{+0.0001}_{-0.0001}$ & $1.1975^{+0.0001}_{-0.0001}$ &  $1.1975^{+0.0001}_{-0.0001}$  & $1.1975^{+0.0001}_{-0.0001}$ & $1.1975^{+0.0001}_{-0.0001}$ & $1.1975^{+0.0001}_{-0.0001}$ \\
        $q$ & $0.88^{+0.08}_{-0.10}$ & $0.87^{+0.08}_{-0.09}$ & $0.88^{+0.08}_{-0.09}$ &   $0.88^{+0.8}_{-0.9}$ &  $0.89^{+0.07}_{-0.08}$ & $0.87^{+0.08}_{-0.09}$ & $0.87^{+0.08}_{-0.09}$ \\
        $a_1$ & $0.02^{+0.02}_{-0.01}$ & $0.02^{+0.02}_{-0.01}$ & $0.02^{+0.02}_{-0.01}$ &  $0.02^{+0.02}_{-0.01}$  &  $0.02^{+0.02}_{-0.01}$ & $0.02^{+0.02}_{-0.01}$ & $0.02^{+0.02}_{-0.01}$ \\
        $a_2$ & $0.02^{+0.02}_{-0.01}$ & $0.02^{+0.02}_{-0.01}$ & $0.02^{+0.02}_{-0.01}$ &  $0.02^{+0.02}_{-0.01}$  &  $0.02^{+0.01}_{-0.02}$ & $0.02^{+0.02}_{-0.02}$ & $0.02^{+0.02}_{-0.01}$ \\
        $\theta_1$ [$\deg$]& $81^{+34}_{-34}$ & $81^{+32}_{-34}$ & $82^{+33}_{-34}$ &   $83^{+33}_{-33}$ &  $74^{+31}_{-30}$ & $79^{+34}_{-32}$&  $80^{+33}_{-32}$ \\
        $\theta_2$ [$\deg$]& $84^{+36}_{-36}$ & $82^{+35}_{-34}$ & $84^{+34}_{-35}$ &  $81^{+34}_{-34}$  & $95^{+34}_{-31}$ & $82^{+34}_{-36}$ & $85^{+37}_{-36}$ \\
        $\phi_{1,2}$ [$\deg$] & $174^{+126}_{-121}$ &$177^{+122}_{-121}$  & $181^{+117}_{-121}$ &  $178^{+118}_{-120}$  & $177^{+125}_{-118}$ & $178^{+117}_{-119}$ & $176^{+120}_{-118}$ \\
        $\phi_{\rm JL}$ [$\deg$] & $178^{+122}_{-122}$ & $174^{+124}_{-119}$ & $179^{+120}_{-125}$ & $176^{+122}_{-120}$ & $180^{+119}_{-120}$ &  $6177^{+120}_{-121}$& $175^{+120}_{-116}$ \\
        $\psi$ [$\deg$] & $88^{+61}_{-68}$ & $88^{+53}_{-73}$ & $89^{+62}_{-60}$ &   $68^{+43}_{-60}$ & $89^{+62}_{-61}$ & $90^{+61}_{-65}$ & $89^{+61}_{-61}$ \\
        $\Lambda_1$ & $250^{+355}_{-172}$ & $274^{+385}_{-187}$ & $270^{+350}_{-184}$ &  $280^{+356}_{-193}$  &  $309^{+335}_{-201}$ & $268^{+333}_{-179}$ & $269^{+336}_{-183}$ \\
        $\Lambda_2$ & $423^{+547}_{-289}$ & $425^{+534}_{-292}$ & $422^{+513}_{-287}$ &  $452^{+533}_{-307}$  & $448^{+498}_{-300}$ & $447^{+537}_{-305}$ & $429^{+532}_{-294}$ \\
    
    \noalign{\smallskip}
    \hline
\end{tabular}
\label{table:results}
\end{table*}
\renewcommand{\arraystretch}{1}

The parameter medians and 16th-84th percentiles are collected in Table \ref{table:results}.  The second column reports the results of the GW-only fit, while the third and fourth column refer to the fit including the broad-band afterglow and GW (GW+AG), and to the complete fit that also includes the centroid (GW+AG+C), respectively.

\begin{figure*}
    \includegraphics[width=0.6\textwidth]{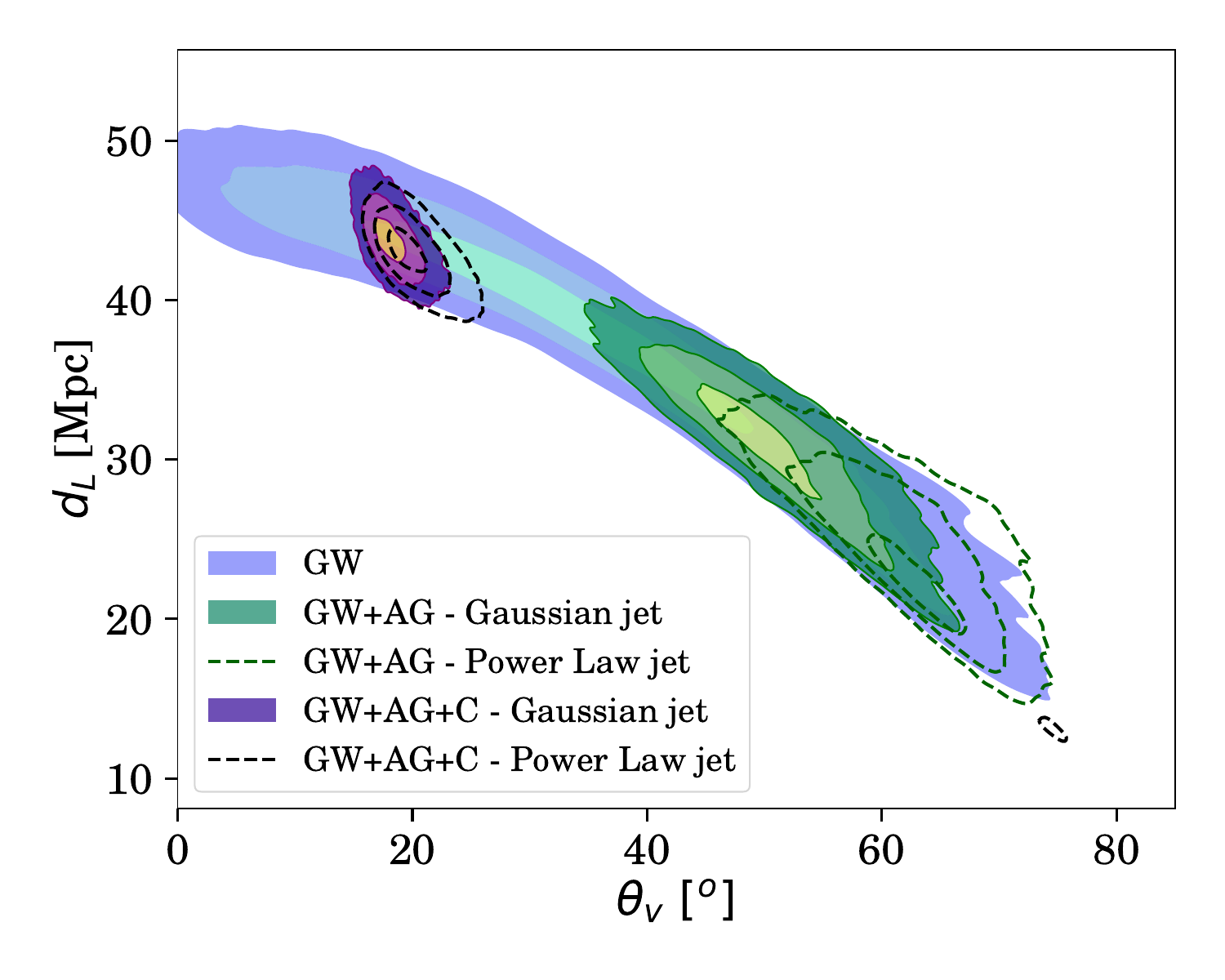}
    \includegraphics[width=0.6\textwidth]{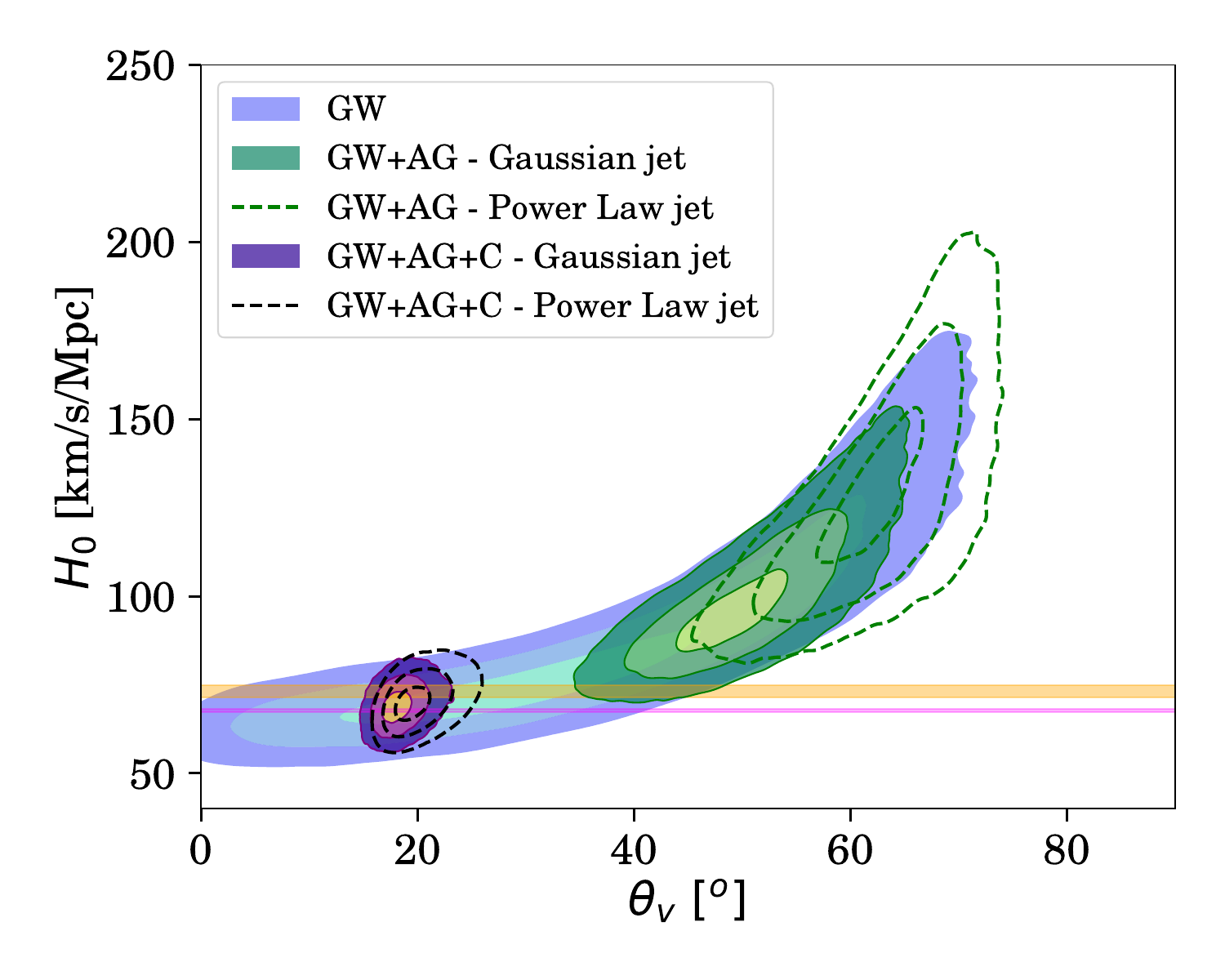}
    \caption{Top panel. Contour plot of the viewing angle and luminosity distance for the GW, GW+AG and GW+AG+C fits. The contours represent the 68\%, 95\%, 99.7\% probability regions. Bottom panel. The same contour plot as above, but switching to $H_0$, instead of $d_L$. The magenta and yellow regions represent the 1 $\sigma$ of the \textit{Planck} and SHoES measurements respectively. The Gaussian jet results are represented with filled contours, while the power law jet with empty contours and dashed lines. }
     \label{Fig:distance}
\end{figure*}

\begin{figure}
    \includegraphics[width=0.45\textwidth]{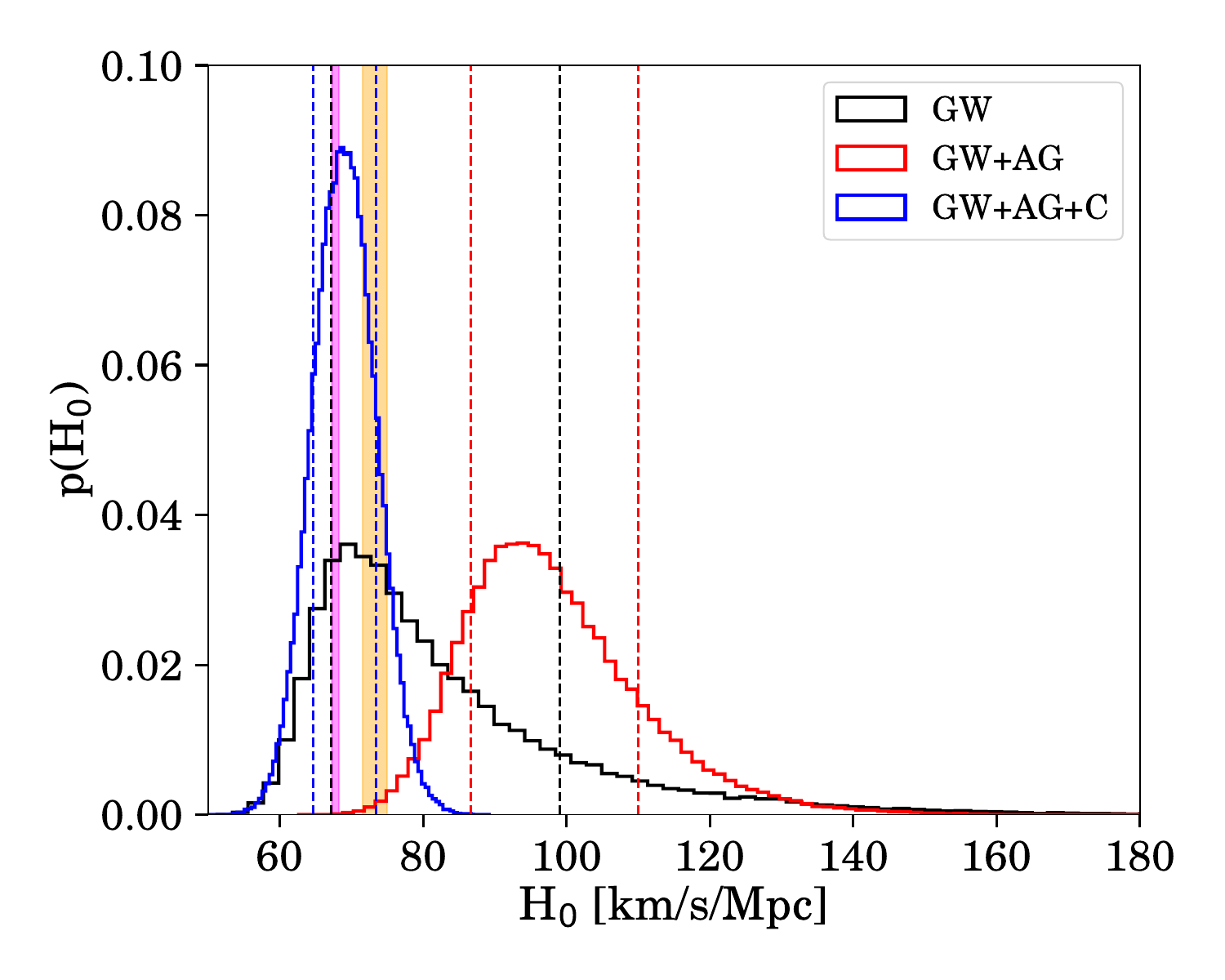}
    \includegraphics[width=0.45\textwidth]{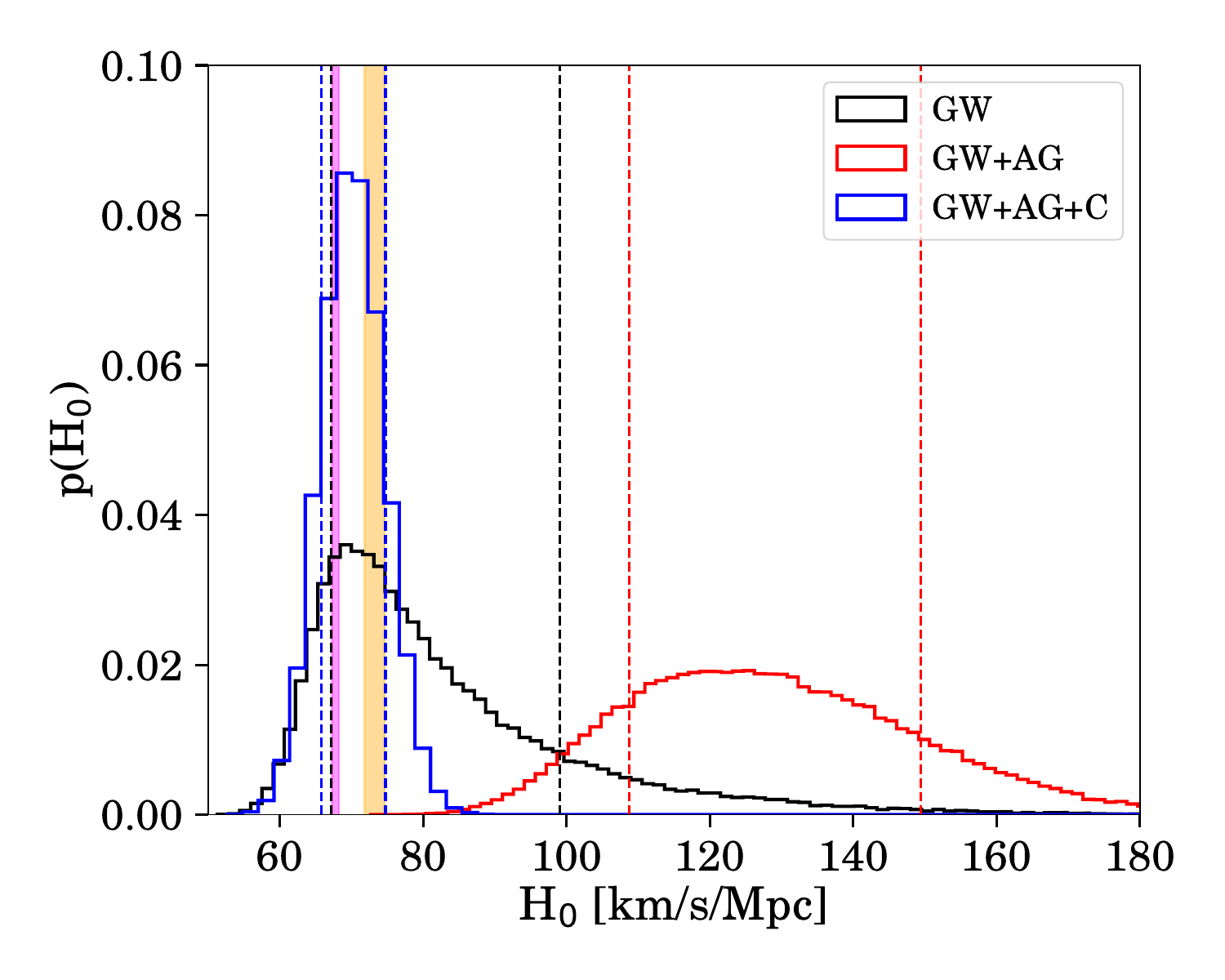}
    \includegraphics[width=0.45\textwidth]{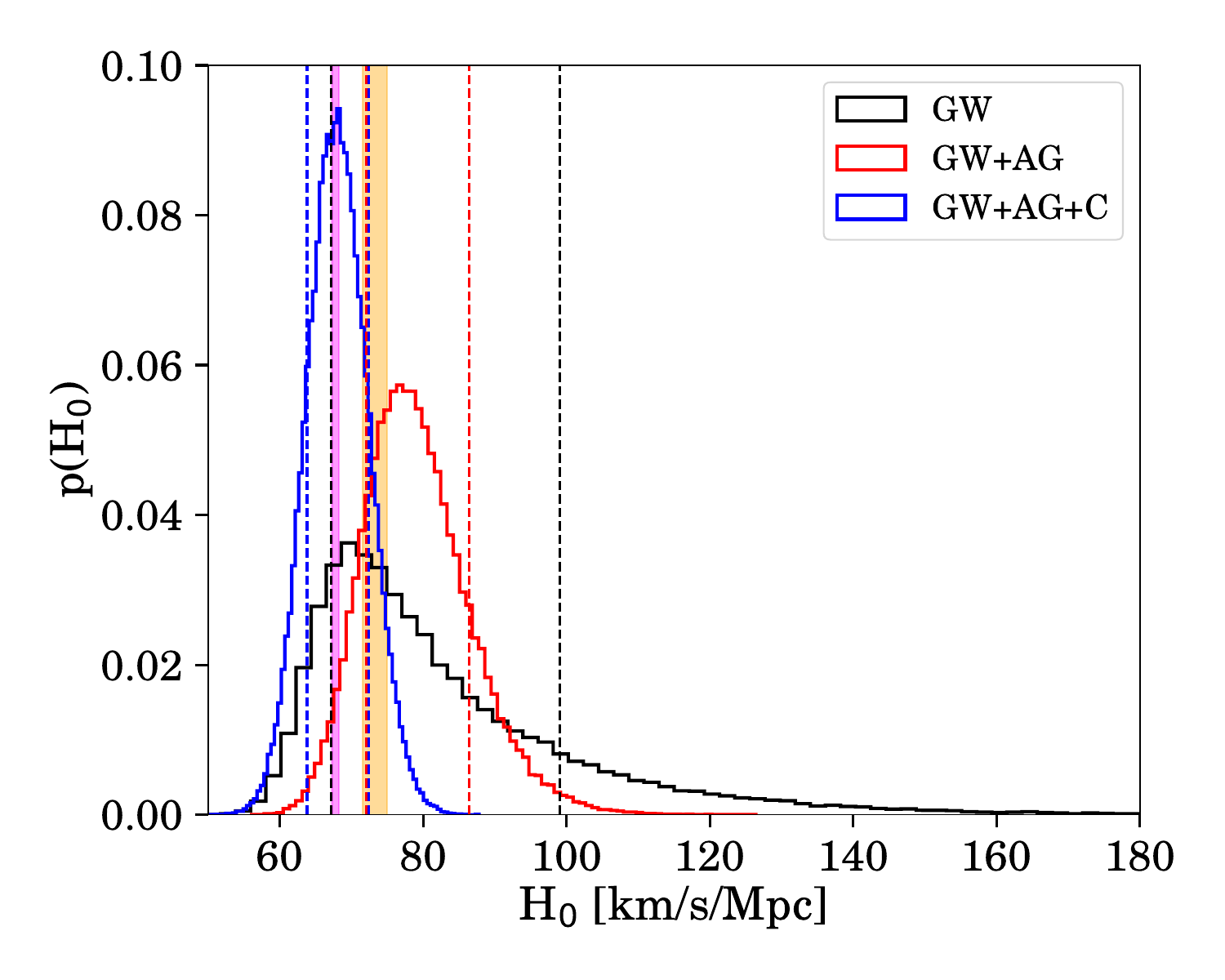}    
    \caption{Histograms of the Hubble constant $H_0$ posterior from the GW-only fit, in black, the GW+AG in red and the GW+AG+C in blue. The vertical dashed lines represent the 16th and 84th percentiles of each distribution. The magenta and yellow shaded regions represent the 1$\sigma$ interval of the \textit{Planck} and SH0ES measurements respectively. Top panel: Gaussian jet. Central panel: power law jet. Bottom panel: Gaussian jet with the addition of a constant component at late times.}
     \label{Fig:H0_hist}
\end{figure}

The results from the GW fit are in agreement with previous works \citet{Abbott2017_dl, Abbott2019_properties, Romero-shaw2020}, the $H_0$ value that we retrieve from the GW-only fit is $H_0 = 77^{+21}_{-10}$\,$\rm km\ s^{-1}Mpc^{-1}$ (median, 16th-84th percentiles), see Fig.\,\ref{Fig:distance}, bottom panel, and Fig.\,\ref{Fig:H0_hist}. As we already pointed out above, one of the main sources of uncertainty in the GW measurement of the inclination and of the distance (and $H_0$) is due to their degeneracy, see the light blue contours in Fig.\,\ref{Fig:distance}, top panel. This means that it is hard to distinguish whether a source is further away with the binary orbit facing Earth (face-on or face-off), or closer but highly inclined (edge-on, \citealt{Usman2019}).
If we assume to have inclinations from 0 to 90 deg (like in our case), $d_L$ is a decreasing function of the inclination (viewing angle, $\theta_v$). Another independent messenger is needed to break this degeneracy, which, in this case, comes from the afterglow.

The afterglow light curve alone, however, is not enough to efficiently break this degeneracy. Including it in the fit, only helps in shrinking the degeneracy region, see green filled contours in Fig.\,\ref{Fig:distance}, top panel, the uncertainty on the viewing angle is reduced by a factor $\sim3$ (from $\theta_{JN}=146^{+16}_{-18} \deg$ to $129.9^{+5.1}_{-5.4} \deg$), the one on the distance by a factor $\sim2$ (from $d_L=39.2^{+5.4}_{-8.6}$ Mpc to $31.3^{+3.0}_{-3.6}$ Mpc). However, this is not an accurate measurement, in fact the medians are on the high-$\theta_v$--low-$d_L$ end of the GW 1$\sigma$ region, leading to quite a low distance (and large viewing angle, $\theta_v = 50.1^{+5.1}_{-5.4} \deg$), which is however within 3$\sigma$ from the generally accepted value of $\sim40$ Mpc. Our $H_0$ value from the GW+AG fit is quite high: we retrieve $H_0 = 96^{+13}_{-10}$\,$\rm km\ s^{-1}Mpc^{-1}$ (median, 16th-84th percentiles), see the green filled contours in Fig.\,\ref{Fig:distance}, bottom panel, and Fig.\,\ref{Fig:H0_hist}, top panel.

As explained in more details in Sections\,\ref{subsec:difference} and \ref{sec:const_flux}, this result is mostly driven by the possible presence of a late time additional component, which can be seen in the top panel of Fig.\,\ref{Fig:lightcurve_gw170817}. The GW+AG model (dotted line) fits very well the light curve, especially the data points at late time. The latter force the model to prefer a high $\theta_v$ with respect to the fit including also the jet centroid motion, GW+AG+C, represented with a solid line. 
Indeed, \citet{Wang2021}, using the same messengers but limiting the light curve data up to $\sim300$ days (when no flux deviation is present yet), retrieve $H_0 = 69.5 \pm 4 \ \rm km\ s^{-1}Mpc^{-1}$, with a $d_L = 43.4\pm1$ Mpc and $\theta_v = 22\pm1$ deg. The jet structure model they use is from 3-dimensional general-relativistic megnetohydrodynamical simulations.
Also \citet{Wang2023} fit the afterglow light curve and include an additional component at late times, a sub-relativistic kilonova outflow. They estimate ${H}_{0}=71.80^{+4.15}_{-4.07} \ \rm km\ s^{-1}Mpc^{-1}$. The kilonova component helps in the fit of the light curve, keeping the viewing angle around 30 deg. Also in this case, they model the jet using hydrodynamic simulations. \citet{Guidorzi2017} get ${H}_{0}=75.5^{+11.6}_{-9.6} \ \rm km\ s^{-1}Mpc^{-1}$, assuming a Top Hat jet and fitting the afterglow data up to 40 days from the merger. The latter is the reason why the $H_0$ uncertainties are larger with respect to more recent works, their $\theta_v$ posterior distribution peaks at $\sim$30 deg. However, the Top Hat jet is not the best choice for GW170817 light curve, as it cannot reproduce the slope before the peak. 

In Section\,\ref{sec:const_flux} we account for the possible late time excess in the GW+AG fit with a constant flux component. In this case, our results are in agreement with the aforementioned works, see the complete analysis in Section\,\ref{sec:const_flux}.  It is also interesting to note that the similar results of these works are obtained using different jet structures. We will come back to this point and explore how $H_0$ changes depending on the jet structure in Section\,\ref{subsec:PL_jet}.
 
\begin{figure*}
    \includegraphics[width=0.4\textwidth]{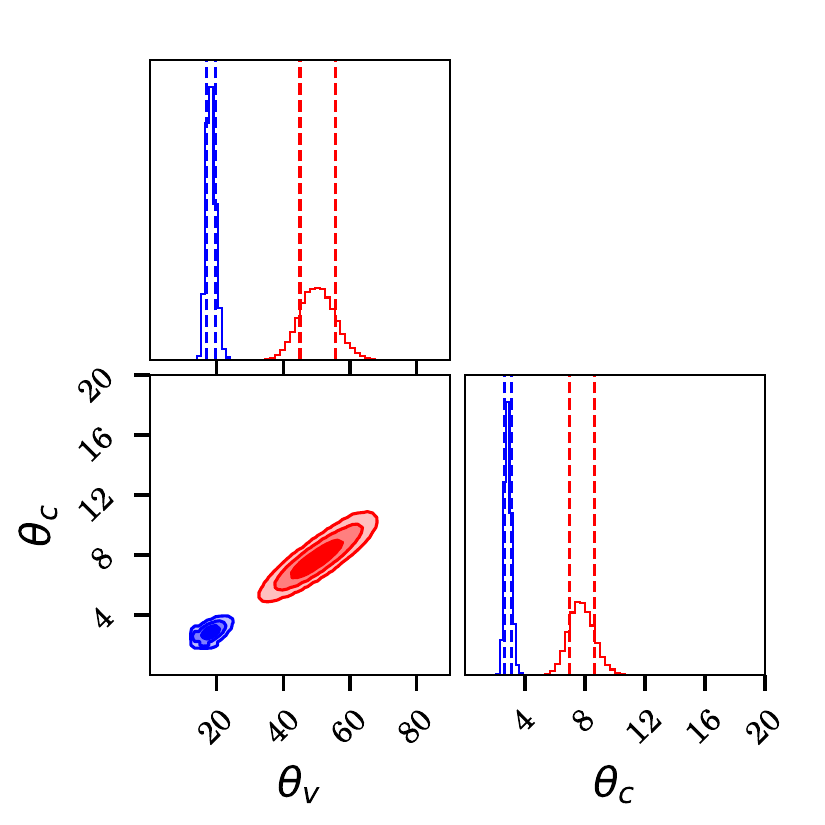}
    \includegraphics[width=0.41\textwidth]{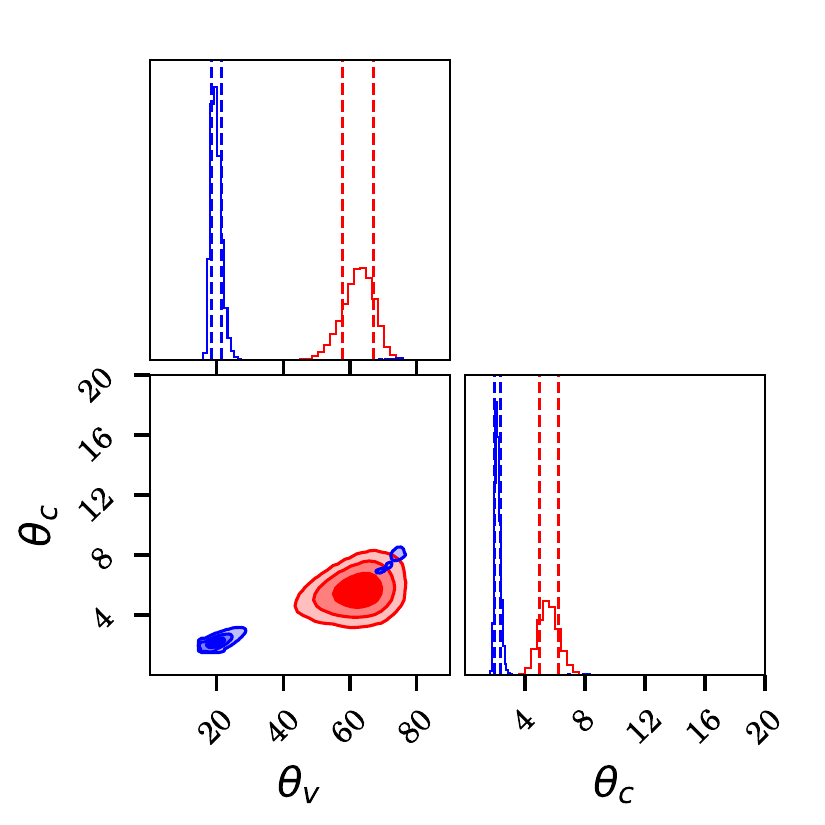}    
    \caption{Contour plots of the viewing angle and jet opening angle. The contours represent the 68\%, 95\%, 99.7\% probabilities. The blue contour lines represent the result from the joint GW+AG+C fit, while the red ones represent the result from the GW+AG fit. Left panel: Gaussian jet. Right panel: power law jet.}
     \label{Fig:angles}
\end{figure*}

From these results, we find that limiting the analysis to GW+AG domains could be subject to possible systematics in the $H_0$ determination due to the detections at late times in the afterglow light curve. This adds up to the degeneracy between $\theta_v$ and $\theta_c$, proper of a Gaussian modelling of the jet, that is evident in the left panel of Fig.\,\ref{Fig:angles}. Here we show the marginalised, 2D posterior probability distributions for the jet opening angle $\theta_c$ and the viewing angle $\theta_v$ in the cases of the the joint GW+AG fit (in red contours). 

For the reasons stated above, in order to break both the $d_L-\theta_v$ GW degeneracy and the $\theta_v-\theta_c$ EM one, we have to include not only the afterglow light curve, but also the centroid motion in the analysis. We note that the sole centroid motion is not enough to break the $d_L-\theta_v$ degeneracy, being itself subjected to some level of degeneracy between these two parameters, see Appendix\,\ref{Sec:Appendix} for a more detailed discussion.

The results for the GW+AG+C fit, using both the afterglow and the centroid, are written in Table\,\ref{table:results}, fourth column. This fit not only shifts the viewing angle to lower values, but also shrinks further the degeneracy between $\theta_v$ and $\theta_c$, see left panel of Fig.\,\ref{Fig:angles}, blue contours. This happens because the relativistic jet motion strongly constrain the viewing angle. In Fig.\,\ref{Fig:lightcurve_gw170817}, bottom panel, the jet positions are well reconstructed by the model, within 1$\sigma$, while in top panel, we see that the GW+AG+C model does not fit well the late time light curve, especially in the X-rays and in the radio bands, unlike the GW+AG fit, recognising it as a possible excess not due to the afterglow emission. We try to account for this adding a constant flux component at late times. The latter helps in fitting that part of the light curve, but results in very similar posteriors to the GW+AG+C fit without it (see the full results in Section\,\ref{sec:const_flux}). This shows that, adding the afterglow centroid motion in the analysis, provides robustness to the fit. The discrepancy in the fit of the light curve at late time when including the jet centroid motion is due to the fits preference for two different viewing angles and types of jet: the GW+AG fit prefers a large $\theta_v$, a broader jet profile and less energy on the jet axis, while the GW+AG+C fit prefers a small $\theta_v$, a highly collimated jet with a large energy on the jet axis and a less dense circumburst medium. We go into detail of this differences in Sec.\,\ref{subsec:difference}.

The $\theta_v$ and $d_L$ posteriors of the GW+AG+C fit are in the low-$\theta_v$--high-$d_L$ GW 1$\sigma$ region of the degeneracy, predicting a distance of $43.7^{+1.4}_{-1.4}$ Mpc and a viewing angle of $18.2^{+1.2}_{-1.5}$ deg. The centroid addition in the fit helps in shrinking the uncertainties in these parameters, which are smaller by a factor of 4-5 for the distance and 8-9 for $\theta_v$ with respect to the GW analysis, breaking their degeneracy, see the purple and yellow filled contours in the top panel of Fig.\,\ref{Fig:distance}. From the GW+AG+C fit we obtain $H_0 = 69.0^{+4.4}_{-4.3}$\,$\rm km\ s^{-1}Mpc^{-1}$ (median, 16th-84th percentiles). It is to be noted that, adding the centroid in the analysis, brings to an about three times more precise $H_0$ measurement than the GW-only standard-siren measurement. The \textit{Planck} estimate of $67.4\pm{0.5}$\,$\rm km\ s^{-1}Mpc^{-1}$ \citep{Planck2020} and the SHoES value of $74.0\pm1.4$\,$\rm km\ s^{-1}Mpc^{-1}$ \citep{Riess2019}, are both within 1$\sigma$, see Fig.\,\ref{Fig:distance}, bottom panel, and Fig.\,\ref{Fig:H0_hist}, top panel. 
This result is in agreement also with other works, like \citet{hotokezaka2018}, who use the posteriors from GW, and fit the afterglow flux and the centroid motion, finding ${H}_{0}=68.9^{+4.7}_{-4.6} \ \rm km\ s^{-1}Mpc^{-1}$. \cite{Palmese2023} use the same model as \citet{Wang2023} (hydrodynamic simulations), and use a prior on the jet break Lorentz factor from the centroid measurements, which acts also on the jet opening angle and on the viewing angle. They find ${H}_{0}=75.46^{+5.34}_{-5.39} \ \rm km\ s^{-1}Mpc^{-1}$. Leaving the Lorentz factor free leads to an opening angle of around 7 deg, which is instead consistent with our GW+AG results for $\theta_c$. 

Our values of $\theta_v$ and $\theta_c$ from the GW+AG+C fit are in agreement with other works that included the centroid motion in their analysis. \cite{Ghirlanda2019} predicts $\theta_c = 3.1\pm1$ deg, with a viewing angle of about 15 deg, \cite{Mooley2018, Mooley2022} an opening angle of $<5$ deg, and a viewing angle $<24$ deg, while \citet{Ren2020} find $\theta_c = 3.1$ deg and $\theta_v = 17.4$ deg.

\subsection{About the difference between GW+AG and GW+AG+C fits}
\label{subsec:difference}

The GW+AG and GW+AG+C produce quite different results, not only regarding the Hubble constant, the luminosity distance and the viewing angle, but also the energetics and microphysics of the jet. 
Indeed, while the GW+AG fit results in a large $\theta_v$, a broader jet profile and less energy on the jet axis, the GW+AG+C fit results in a small $\theta_v$, a highly collimated jet with a large energy on the jet axis and a less dense circumburst medium. The viewing angle values are about 5$\sigma$ away, which is quite singular, considering that the event is the same.
This, as we stated above, is due to the light curve data points at late times, which are well captured by the GW+AG fit, but not by the GW+AG+C fit, see Fig.\,\ref{Fig:lightcurve_gw170817}. In particular, when including the afterglow centroid motion in the fit, the latter prevails over the light curve data points at late times, resulting in a low viewing angle and a fit of the light curve that, at late times, shows deviations from the flux observations.

In the GW+AG+C, the centroid motion is able to constrain very well $\theta_v$ to $18.2^{+1.2}_{-1.5}$ deg, which then translates into a constraint also on $\theta_c = 2.85^{+0.24}_{-0.20}$ deg. This happens because of the degeneracy between the two angles, proper of the Gaussian jet light curve (see Fig.\,\ref{Fig:angles}). In fact, its rising slope depends on their ratio, which, in this fit, is about 6.4. In the GW+AG, instead, there are no constraints on $\theta_v$ or $\theta_c$ individually, but just on their ratio, from the rising slope of the light curve \citep[see also][]{Ryan2020, Nakar2021}. This still leads to the same ratio of about 6.5, but $\theta_v=50.1^{+5.1}_{-5.4}$ deg and $\theta_c=7.73^{+0.86}_{-0.80}$ deg, about 5 sigma away from the GW+AG+C case.

The GW+AG fit, not being constrained by the centroid data set, is free to account for the mild decay of the light curve at late times by anticipating the non-relativistic phase. In particular, we estimate the non-relativistic time \citep{Ryan2020} to be $t_{NR} = 880^{+290}_{-210}$ days (GW+AG), with respect to $t_{NR} = 13000^{+2700}_{-2400}$ days (GW+AG+C). Therefore, at late times, according to the parameters of the GW+AG fit, the jet is non-relativistic. 
The anticipation of the relativistic phase is obtained mainly by acting on the $E_0,n_0$ parameters. However, the one order of magnitude lower energy and two orders of magnitude higher circumburst density would shift the flux at low values and the break at earlier times, since $t_b \propto (E_0/n_0)^{1/3} (\theta_v + 1.24\theta_c)^{8/3}$ \citep{Ryan2020}. This is balanced by the fit with higher values of $\theta_c$ and $\theta_v$, in order to bring back the jet break (the peak) at about 130 days, and to adjust the rising slope of the light curve. This influences also the early decreasing slope (before the non-relativistic phase), as a higher $\theta_c$ (wider jet) provides a larger surface area, so the jet is brighter and the flux is higher. The parameters $d_L$, $\epsilon_e$, $\epsilon_B$ have mainly the role of shifting the flux. The $p$ parameter stays the same in the two fits, as it is constrained by the spectrum. $\theta_W$ is unconstrained in both fits, however it is better constrained in the GW+AG fit, mainly because $\theta_c$ is larger, and, being a Gaussian jet, $\theta_w$ has to be lower than $\theta_c$.


In other words, the good fit in the GW+AG case is provided by a combined effect of the high $\theta_c$ (in the decreasing slope right after the peak) and the anticipation of the non-relativistic phase (in the slope at late times).



\subsection{Changing the structure of the jet}
\label{subsec:PL_jet}

\begin{figure}
    \includegraphics[width=0.5\textwidth]{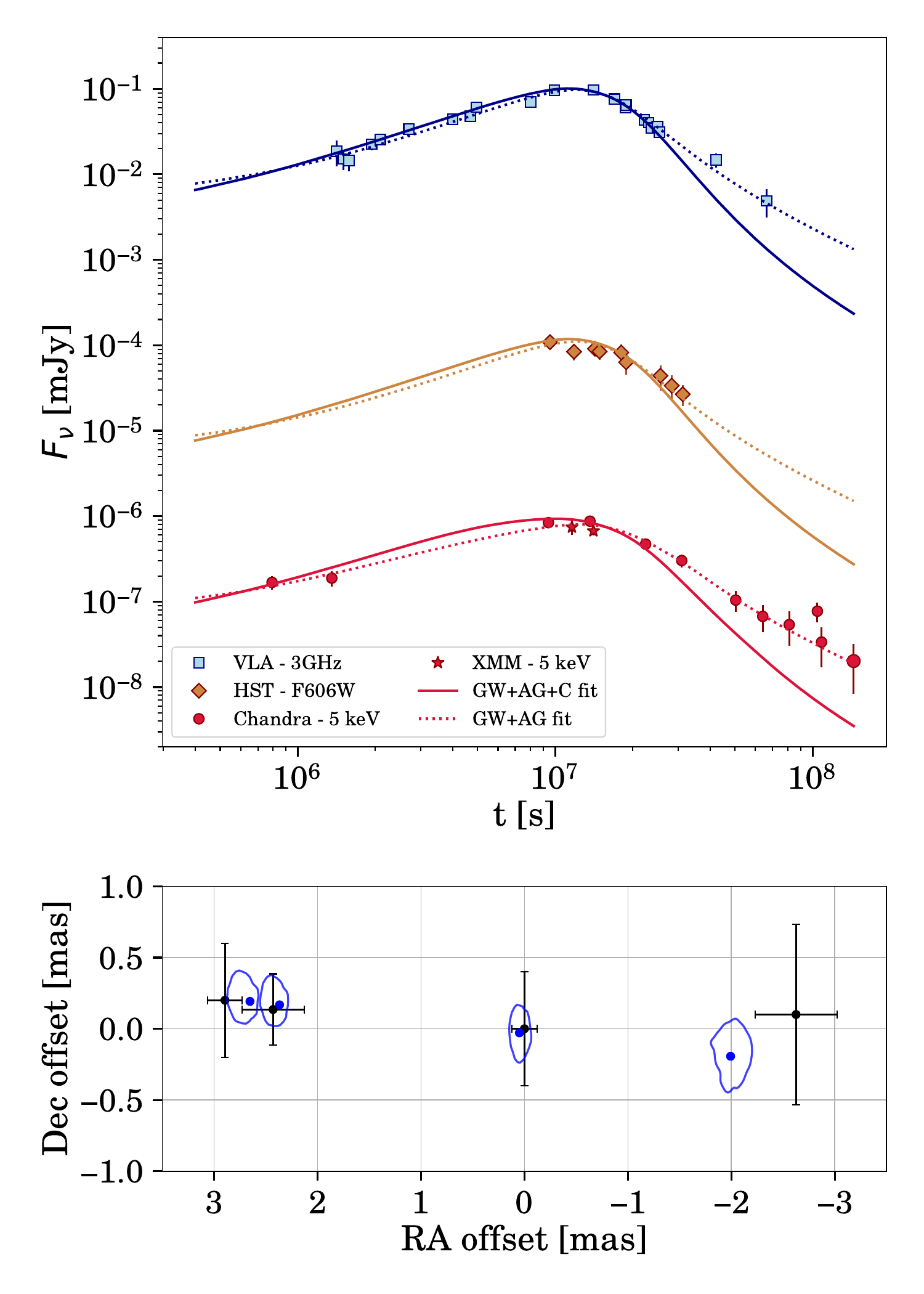}
    \caption{Same as Fig.\,\ref{Fig:lightcurve_gw170817}, but assuming a power law structure for the jet.}
     \label{Fig:lightcurve_gw170817_pl}
\end{figure}

In the case of a power law jet, the degeneracy between $\theta_v$ and $\theta_c$ is not as strong as for the Gaussian geometry, the rising phase slope is a function of $b$, $\theta_v$ and $\theta_c$ (see Eq.(33) of \citealt{Ryan2020}). This can be seen in the right panel of Fig.\,\ref{Fig:angles}, where the GW+AG fit is represented in red contours. The GW+AG and GW+AG+C are written in the fifth and sixth column of Table\,\ref{table:results}, while the fits of the afterglow light curve and centroid motion are in Fig.\,\ref{Fig:lightcurve_gw170817_pl}. Also for this jet structure, the GW+AG and GW+AG+C produce quite different results, and the reasoning in Section\,\ref{subsec:difference} is still valid. The majority of the parameters from the GW+AG+C and GW+AG fits, assuming a power law model, are in agreement within $1\sigma$ with the Gaussian jet model, this is probably due to the fact that, at early times, the afterglow light curve rises, so $b$ has to be large. At the same time, the larger is $b$, the more the Gaussian and power law structures are similar. For example, in the case of the GW+AG+C fit, for $b=10.8$, the $E(\theta)$ of a Gaussian and power law structures are very similar within $\sim3\theta_c$, after which the decay is shallower for the power law structure.  

The GW+AG fit produces a larger $\theta_v = 62.7^{+5.0}_{-4.3}$ deg, a smaller $\theta_c = 5.57^{+0.69}_{-0.62}$ deg and smaller $d_L = 23.7^{+3.8}_{-3.4}$ Mpc, than the Gaussian jet, these parameters are, however, in agreement within 2$\sigma$ with the latter. The 2D posterior for $\theta_v$ and $d_L$ are represented in Fig.\,\ref{Fig:distance}, top panel, in green dashed contours. The microphysics and the energetics are in agreement within $1\sigma$ with the Gaussian jet results. 

In the GW+AG+C fit, the parameters are in agreement within $1\sigma$ with the Gaussian jet model, except for $\theta_c = 2.18^{+0.20}_{-0.16}$ deg, which is within $2\sigma$. The $\theta_v$ and $\theta_c$ 2D posteriors for the power law jet are represented in the right panel of Fig.\,\ref{Fig:angles}, we can see that at 3$\sigma$ there are samples with large viewing angles, usually preferred instead from the GW+AG fits. The results for $\theta_v$ is $19.8^{+1.3}_{-1.8}$ deg and for $d_L = 43.0^{+1.4}_{-1.4}$ Mpc.
The 2D posterior distributions of $\theta_v$ and $d_L$ are in Fig.\,\ref{Fig:distance}, top panel, in black dashed contours, almost superimposed to the Gaussian jet results (purple and yellow coloured contours). Also in this case there is a small region of the parameter space at 3$\sigma$ at large $\theta_v$ and small $d_L$, which gets cancelled when estimating $H_0$, bottom panel.

The $H_0$ values retrieved in these fits are $70.2^{+4.6}_{-4.4}$ $\rm km\ s^{-1}Mpc^{-1}$ for the GW+AG+C fit, and $127^{+22}_{-19}$ $\rm km\ s^{-1}Mpc^{-1}$ (medians, 16th-84th percentiles) for the GW+AG fit, the $H_0$ posteriors are represented in the central panel of Fig.\,\ref{Fig:H0_hist}, in blue and red respectively. 
For this event, if we use the complete data set (the most robust fit), the change in jet model does not significantly influence $H_0$, the power law jet predicts an $H_0$ larger than 1.2 $\rm km\ s^{-1}Mpc^{-1}$, which is a $2\%$ difference, with respect to a Gaussian jet, but still in agreement within the uncertainties. In the GW+AG there is a $30\%$ difference, but the two $H_0$s are compatible within $1\sigma$.

In order to assess if the unknown jet structure leads to systematics in the estimation of $H_0$, we simulate an afterglow light curve and centroid movement using a Gaussian jet, then we fit them twice, assuming a Gaussian and a power law structure. To keep this simulation as similar to GW170817 as possible, we keep the GW170817 detection times, errors and frequencies for the afterglow light curve and centroid motion, but we adopt fluxes and positions predicted by the model, with a Gaussian variation. We simulate the EM data sets assuming a Gaussian jet and the parameters in Table\,\ref{table:results}, medians in the fourth column. In this way, we do not include the excess in the flux at late times, which we are not interested in, as we are focusing on the influence of the jet structure.
These EM data sets are then fitted with GW two times, one assuming a Gaussian jet, and the other assuming a power law jet. 

\begin{figure}
    \includegraphics[width=0.5\textwidth]{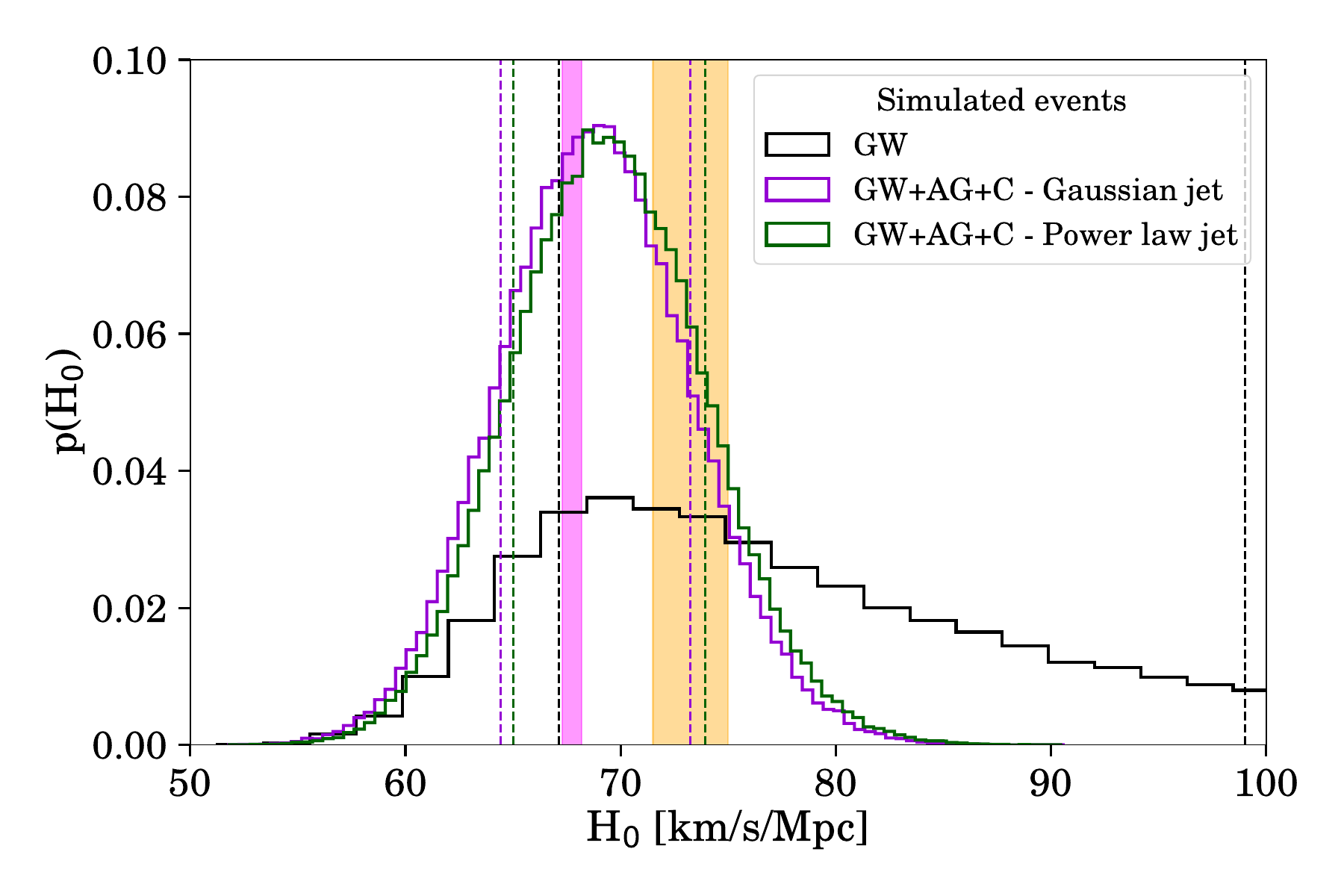}
    \caption{Histogram of the $H_0$ posterior distribution for a simulated event. The GW-only fit is represented in black (same distribution as Fig.\,\ref{Fig:H0_hist}), the GW+AG+C assuming a Gaussian jet in the fit is represented in violet, while the GW+AG+C assuming a power law jet is represented in green. The vertical dashed lines represent the 16th and 84th percentiles of each distribution. The magenta and yellow shaded regions represent the 1$\sigma$ interval of the \textit{Planck} and SH0ES measurements respectively. }
     \label{Fig:H0_hist_simul}
\end{figure}

For both jet structures, we retrieve the parameters of the GW, the energetics and microphysics in agreeement within 1$\sigma$ with the median values in Table\,\ref{table:results}, fourth column. Focusing on the distance and the geometry of the system, assuming a Gaussian jet, we retrieve $\theta_v = 19.3^{+1.5}_{-1.7}$ deg, $\theta_c = 3.01^{+0.28}_{-0.25}$ deg and $d_L = 43.8^{+1.5}_{-1.7}$ Mpc, while for a power law jet $\theta_v = 20.2^{+1.6}_{-1.8}$ deg, $\theta_c = 2.40^{+0.24}_{-0.21}$ deg and $d_L = 43.6^{+1.5}_{-1.5}$ Mpc. As for the case of GW170817, the power law jet tends to give a slightly higher (lower) viewing angle (jet opening angle), which is in agreement within 2$\sigma$ with the simulated values. This, however, does not influence much the luminosity distance. The $H_0$ posteriors that we retrieve from these fits are represented in Fig.\,\ref{Fig:H0_hist_simul}, with purple (Gaussian jet fit) and green (power law jet fit) colors. It seems that the Hubble constant, as $d_L$, is not influenced by the different structure, resulting in $H_0 = 68.8^{+4.5}_{-4.4}$\,$\rm km\ s^{-1}Mpc^{-1}$ for a Gaussian jet and $H_0 = 69.4^{+4.5}_{-4.4}$\,$\rm km\ s^{-1}Mpc^{-1}$ for a power law jet (medians, 16th-84th percentiles). This is a less than $1\%$ difference, which is well inside the $1\sigma$ range, nonetheless is also at the same level of the \textit{Planck} uncertainty on $H_0$. For this reason, in the future, with a larger number of events, it could be important to assess if this, at the moment negligible difference, is just a statistical fluctuation or a real fluctuation due to the changing jet structure. 

\subsection{Adding a constant component in the flux at late times}
\label{sec:const_flux}

In the case of GW170817, the high viewing angle preference mainly arise at late times, where there is a flux excess. This is either due to some missing emission at late times in the jet model itself, or due to a new component becoming visible, like a kilonova afterglow or the emission from a long-lived pulsar, in the former case we expect to see a rising flux in future observations, in the latter a constant flux \citep{Troja2020, Hajela2019, Piro2018}. If a flux additive component is included in the fit, indeed the jet viewing angle slightly decreases, see for example \citet{troja2021, Balasubramanian2021, Hajela2022, Wang2023, Ryan2023}.

We fit the same data set in Fig.\,\ref{Fig:lightcurve_gw170817}, adding a constant flux component of the type $F_{\nu} = F_{\nu, \rm agpy} + 10^{c}$, where $F_{\nu, \rm agpy}$ is the flux predicted by \afterglowpy and $c$ is a parameter in the fit. This is done only at late times and at all frequencies. The $c$ parameter has three possible values, depending on the frequency: $c_{\rm radio}$, with a uniform prior in [-3.5,-2], $c_{\rm optical}$ with a uniform prior in [-5.5, -4.5] and $c_{\rm X-rays}$, with uniform prior in [-8, -7].

\begin{figure}
    \includegraphics[width=0.5\textwidth]{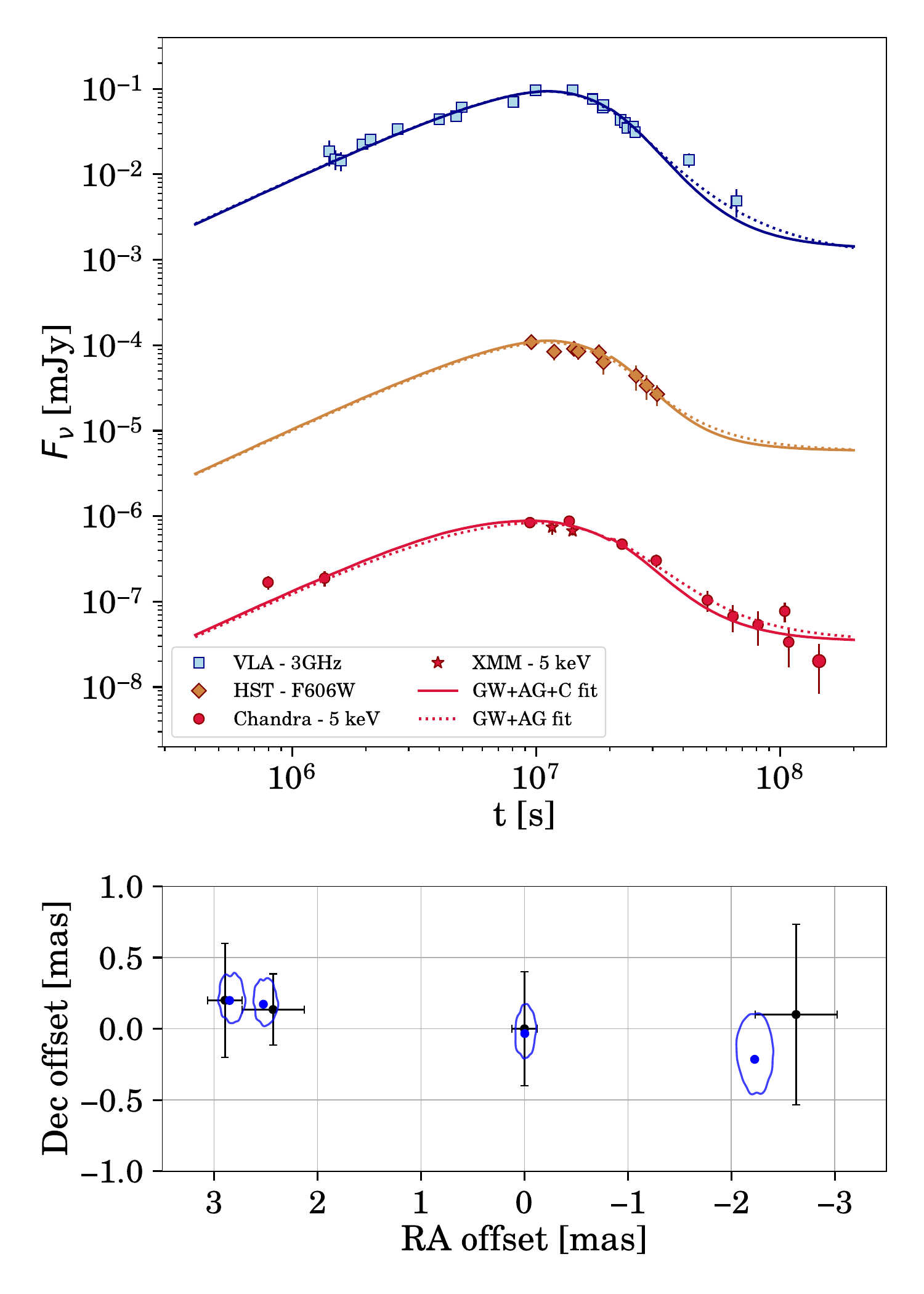}
    \caption{Same as Fig.\,\ref{Fig:lightcurve_gw170817}, but including an additional constant flux component in the model at late times.}
     \label{Fig:lightcurve_const}
\end{figure}

The results of the GW+AG+C and GW+AG fit are written in Table\,\ref{table:results}, last two columns, while the fit of the broad-band afterglow light curve and centroid motion are in Fig.\,\ref{Fig:lightcurve_const}.

\begin{figure}
    \includegraphics[width=0.5\textwidth]{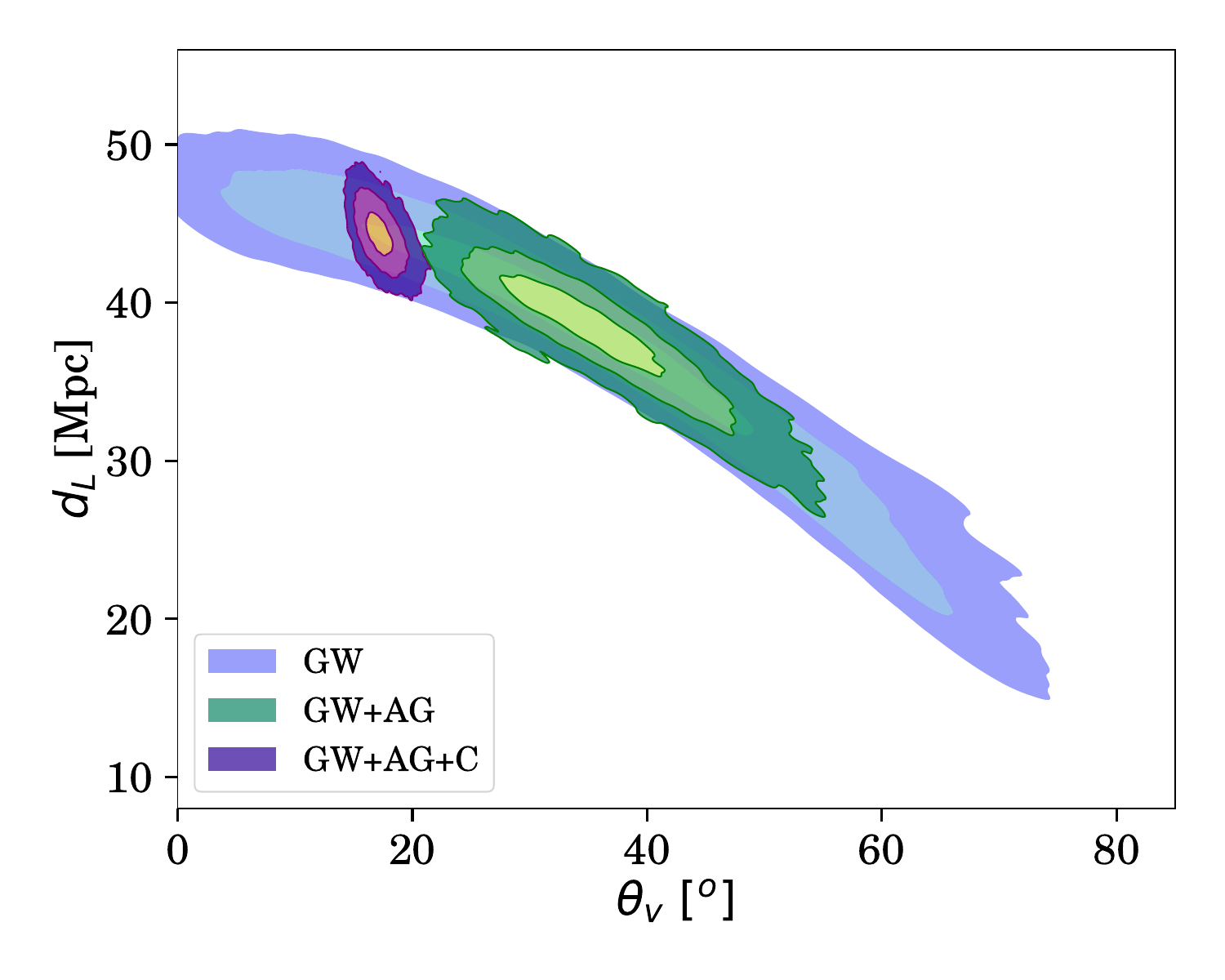}
    \includegraphics[width=0.5\textwidth]{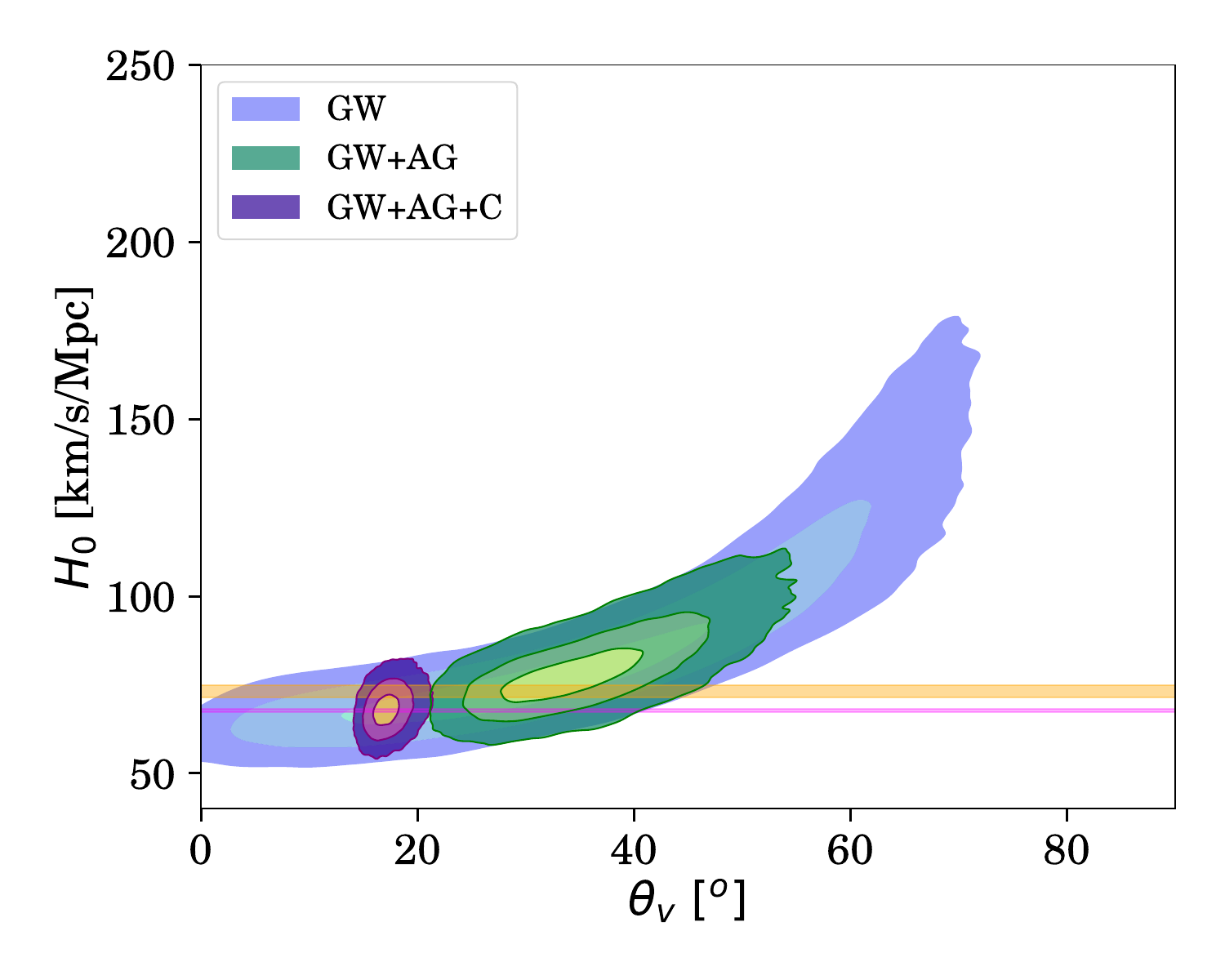}
    \caption{Same as Fig.\,\ref{Fig:distance}, but using a Gaussian jet with the addition of a constant flux component at late times.}
     \label{Fig:distance_const}
\end{figure}

This model can well fit the afterglow light curve and centroid motion, in both cases. The parameters values from the GW+AG+C fit are in agreement within 1$\sigma$ with the ones from the simple Gaussian jet model (Table\,\ref{table:results}, fourth column). In the GW+AG+C the viewing angle $\theta_v=17.2^{+1.1}_{-1.2}$ deg is lower with respect to the simple Gaussian jet model, so the distance $d_L = 44.3^{+1.4}_{-1.3}$ Mpc is slightly larger (see, for example, Fig.\,\ref{Fig:distance_const}, top panel). The viewing angle and the jet opening angle are better constrained, but the error on the distance is unvaried with respect to the previous analysis. This fit leads to an $H_0$ of $68.0^{+4.4}_{-4.2}$ $\rm km\ s^{-1}Mpc^{-1}$ (see bottom panel of Fig.\,\ref{Fig:H0_hist} and bottom panel of Fig.\,\ref{Fig:distance_const}), which is in agreement with the value from the GW+AG+C fit with a simple Gaussian jet. 

The inclusion of a constant component that accounts for the late-time behaviour does not significantly influence the parameter posteriors with respect to the model without it, so we can say that the GW+AG+C fit and model are robust.

In the case of the GW+AG fit, the addition of the constant component brings some improvements in the results. The jet parameters are compatible at most within 2$\sigma$ with the GW+AG+C (with constant) fit, except for $\theta_c = 5.37^{+0.97}_{-0.87}$ deg and $\theta_v= 35.2^{+5.7}_{-6.2}$ deg, which are within 3$\sigma$. Thanks to the inclusion of the constant component at late times, the viewing angle decreases with respect to the GW+AG fit with the simple Gaussian jet model, and the error on the distance is about 2 times better that the GW fit, cutting part of the tails of the $\theta_v-d_L$ degeneracy,  see Fig.\,\ref{Fig:distance_const}, top panel. Indeed, the Hubble constant value that we retrieve is $78.5^{+7.9}_{-6.4}$ $\rm km\ s^{-1}Mpc^{-1}$, see the bottom panels of Fig.\,\ref{Fig:H0_hist} and Fig.\,\ref{Fig:distance_const}, that is compatible within 1$\sigma$ with the GW+AG+C fit, including a constant component. Moreover, this result is compatible within 1$\sigma$ with \citet{Guidorzi2017, Wang2021, Wang2023}.

\subsection{Prospects for jet centroid observations}

Using GW, afterglow light curve and centroid motion, leads to $H_0 = 69.0^{+4.4}_{-4.3}$\,$\rm km\ s^{-1}Mpc^{-1}$. However, the precision of this measure is not at the level of SH0ES or \textit{Planck}, in order to reach these, we would need at least $\sim10$ events and $\sim60$ events respectively. In this Section, we estimate the likelihood that a new GW event, followed by the detection of the afterglow light curve and the measurement of the afterglow centroid motion, is seen in the next GW Observing runs O4 and O5.

From the GW simulations of \citet{Petrov2022}, we generate the EM counterparts of more than a thousand binary neutron star events detectable in O4 \citep{singerO42021} and in O5 \citep{singerO52021}, assuming a Gaussian jet. 
Each GW event is characterized by an inclination and a luminosity distance, which we use to generate their afterglow light curve and centroid motion. For inclinations larger than 90 deg, we convert them in EM viewing angles as explained at the end of Section\,\ref{subsubsec:gw_model} and in Eq.\,(1) of \citealt{Gianfagna2023}.
We assume all the other parameters to be the same as GW170817 (see Table\,\ref{table:results}, fourth column). Moreover, we assume that all the events are well localized and easy to be followed up by the radio telescopes. This will lead to very optimistic rates.
We adopt VLBI as the reference radio facility, both for O4 and O5, so we assume a sensitivity in the radio band of 24$\rm \mu Jy$ (the observations of GW170817 afterglow centroid motion reached an RMS of about 8$\rm \mu Jy$), and a resolution of 1.5 mas \citep{Ghirlanda2019}. These performances can be achieved also, for example, with the European VLBI Network (EVN)\footnote{\href{https://www.evlbi.org/}{EVN website}}.

The centroid data set is composed of the same detection times of GW170817, but we adopt fluxes and positions predicted by the model. We assume that the afterglow centroid motion is visible if the offset between two data points is above the assumed resolution. Regarding the afterglow light curve, we define an event as detectable if its afterglow peak is above the sensitivity.

\begin{figure}
    \includegraphics[width=0.5\textwidth]{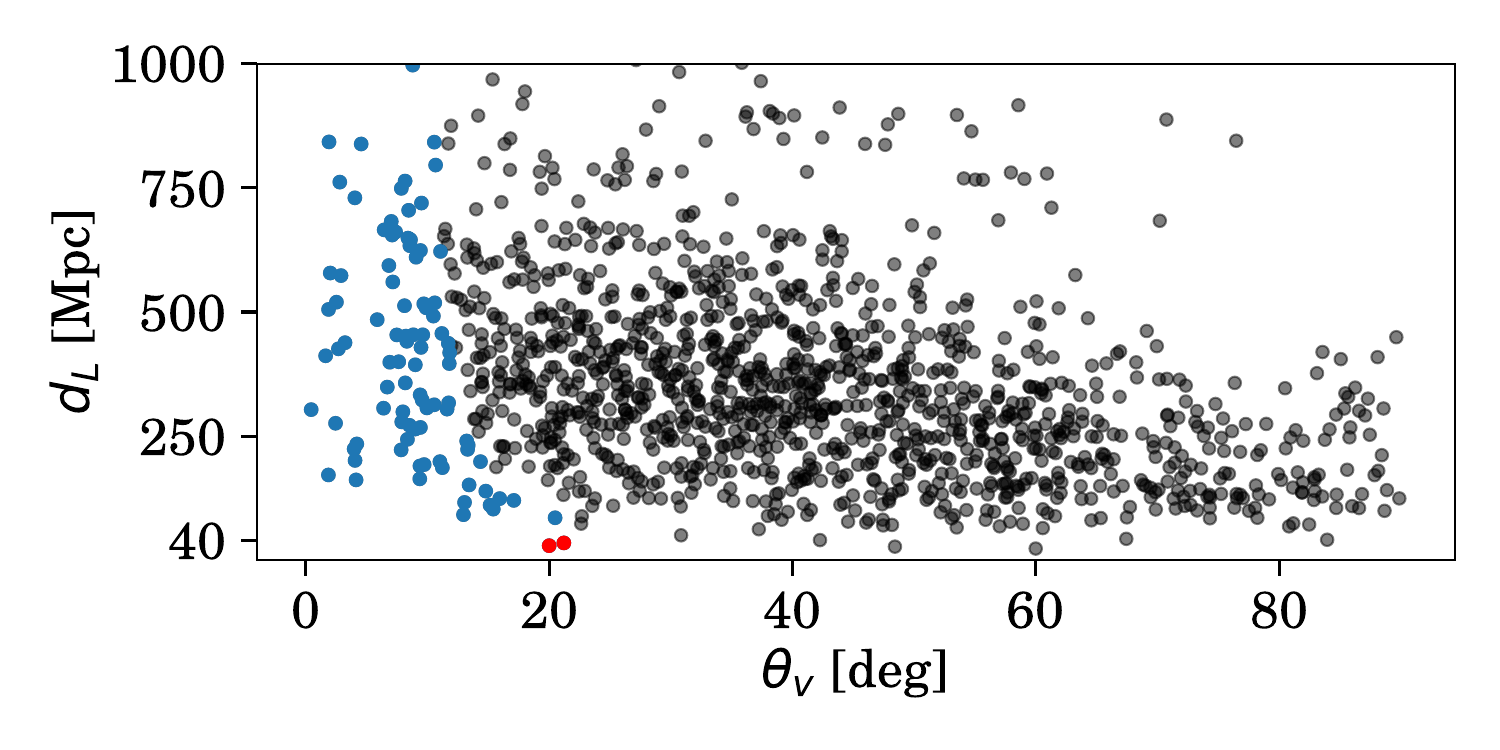}  
    \caption{The dots represent GW events simulated by \citep{Petrov2022}, in the case of the O4 run \citep{singerO42021}. Depending on their $\theta_v$ and $d_L$, we highlight in blue the ones that have a detectable afterglow counterpart in the radio band and in red the ones that have also a detectable afterglow centroid motion.}
     \label{Fig:sims}
\end{figure}

In the case of O4 (operating from 2023 to 2025), the GW rate of events is $34^{+78}_{-25} \rm yr^{-1}$ \citep{Petrov2022}. We find that 7$\%$ of the total have detectable flux, resulting in a rate of $2.4^{+5.5}_{-1.8} \rm yr^{-1}$ (for the whole sky). Regarding jet centroid observations,  we find that only 0.13$\%$ of events has a detectable afterglow flux and centroid, see red dots in Fig.\,\ref{Fig:sims}, in agreement with \citet{Mastrogiovanni2021}. This translate into a rate of $0.05^{+0.11}_{-0.03} \rm yr^{-1}$, therefore it is very unlikely that the jet centroid will be measured again during O4.

In the case of the O5 run, which is due after 2027, the predicted GW rate is $190^{+410}_{-130} \rm yr^{-1}$ . 
We find that 6$\%$ have a peak flux above the sensitivity, resulting in a rate of $11^{+25}_{-8} \rm yr^{-1}$. The jet centroid motion is visible in $0.09\%$ of the cases leading to a rate of $0.17^{+0.36}_{-0.12} \rm yr^{-1}$.
The latter is still a very low rate, with a slightly lower event fraction than O4, due to the fact that O5 will probe larger distances, which very unlikely will have a detectable afterglow centroid motion. The event rate for the GW, afterglow light curve and centroid motion is slightly larger than O4, despite the same number of events at distances lower than 100 Mpc ($\sim 1 \rm yr^{-1}$ both for O4 and O5). For this reason, we can say that the rate fluctuation is just due to the small number of events. 

As is shown in Fig.\,\ref{Fig:sims}, at large distances we mainly see on-axis or almost on-axis events (with a small $\theta_v$), these events will not have a visible jet motion, as the observer is within (or just outside) the jet's opening angle. This results in a small or null offset, which is hardly detected with sensitivities of the order of the mas.
However, if the jet has a large viewing angle, the peak of the afterglow will be at low fluxes, not reaching the VLBI sensitivity. Indeed, for the O4 run, the events that have a coincident detectable GW, afterglow light curve and centroid motion are very similar to GW170817 (at small distances and with $\theta_v\sim20$ deg, red dots in Fig.\,\ref{Fig:sims}).

\section{Conclusions}
\label{sec:conclusions}

The estimation of the Hubble constant $H_0$ exploiting GW, also known as standard sirens method, is a very powerful tool to try to solve the Hubble tension. However, its main issue is the degeneracy between the viewing angle and the luminosity distance of the event, which precludes reaching the level of precision of \textit{Planck} and SH0ES.
In this work, we use this method to estimate $H_0$, with additional constraints that help in breaking this degeneracy. 
Using Bayesian analysis, we fit simultaneously the EM and GW domains for the event GW170817. The electromagnetic data set includes the broad-band afterglow and the centroid motion of the relativistic jet from HST and VLBI observations. From here, we estimate the Hubble constant and we test its robustness depending on the data set used, on the assumed structure of the jet and on the presence of a possible late time flux excess in the afterglow light curve. 

A GW-only fit leads to an Hubble constant value of $H_0 = 77^{+21}_{-10}$\,$\rm km\ s^{-1}Mpc^{-1}$ (median, 16th-84th percentiles). The almost 20$\%$ error is due to the degeneracy stated above. The latter can be broken exploiting independent EM messengers, like the afterglow light curve and centroid motion, at least in the case of GW170817. 

In GW+AG analysis, we join the GW and the afterglow light curve. This fit reduces the $\theta_v-d_L$ degeneracy, but gives $H_0 = 96^{+13}_{-10}$\,$\rm km\ s^{-1}Mpc^{-1}$. This high value follows from the low value of distance ($d_L = 31.3^{+3.0}_{-3.6}$ Mpc) and high value of viewing angle ($\theta_v = 50.1^{+5.1}_{-5.4} \deg$). This behaviour is caused by a possible late time excess in the afterglow flux, these data points are well modelled and are driving the result of the fit. Therefore, for the specific case of GW170817, using only the afterglow as EM counterpart is not enough to get a reliable measurement of $H_0$.

The GW+AG+C fit, instead, joining the GW, the afterglow light curve and the centroid motion, breaks the $\theta_v-d_L$ degeneracy and results in $H_0 = 69.0^{+4.4}_{-4.3}$\,$\rm km\ s^{-1}Mpc^{-1}$, which is in agreement with other estimations of this parameter using GW170817 and is about 3 times more precise than the GW-only $H_0$ measurement. This is because of the very strong constraint on the viewing angle given by the afterglow centroid data set. As a consequence, the latter model does not fit well (even if the residuals are $\leq3.5\sigma$) the late time flux data points. The viewing angle is $\theta_v = 18.2^{+1.2}_{-1.5} \deg$ and the distance is $d_L = 43.7^{+1.4}_{-1.4}$ Mpc. Thus, in the GW+AG+C fit a small value of $\theta_v$, and consequently a highly collimated jet and a large energy on the jet axis, is preferred. In the GW+AG fit a large $\theta_v$, with a broader profile and less energy on the jet axis, is preferred instead.

The possible excess in the afterglow light curve at late times can be explained as either something missing in the jet model, or as a new emission becoming visible \citep{Troja2020, Hajela2019, Piro2018}. In either cases, adding a constant flux component to the GW+AG+C model at late times leads to posterior probabilities that are in agreement within $1\sigma$ with the fit without this constant component ($H_0 = 68.0^{+4.4}_{-4.2}$ $\rm km\ s^{-1}Mpc^{-1}$), but helps in better fit the late times data. This shows that the model and the GW+AG+C results are robust. Instead, adding this constant flux component to the fit of GW+AG leads to more acceptable values of viewing angle, luminosity distance and Hubble constant: $\theta_v= 35.2^{+5.7}_{-6.2}$ deg, $d_L = 38.6^{+2.5}_{-3.0}$ Mpc and $H_0 = 78.5^{+7.9}_{-6.4}$ $\rm km\ s^{-1}Mpc^{-1}$. The latter is compatible within 1$\sigma$ with the GW+AG+C fit. 

Finally, it seems that the Hubble constant is not influenced by the assumption on the structure of the jet (either Gaussian or power law), at the present level of precision. 

We also note that other systematic uncertainties in the Hubble constant estimation can arise from the estimation of the peculiar velocity of the host galaxy \citep{Hjorth2017, Nicolaou2020}. The latter, in this work (see Eq.\,(\ref{eq:vH})), is included in the Hubble flow velocity \citep{Abbott2017_dl}. 
In our analysis, a shift in the NGC4993 peculiar velocity of $\sim140 \rm km/s$ leads to a shift in $H_0$ of $\sim 4 \rm km\ s^{-1}Mpc^{-1}$, in agreement with \citet{Nicolaou2020}. This is still inside the $H_0$ precision reached in this work, but, for a larger number of events, will become one of the main sources of uncertainty.

The best $H_0$ precision reached with this method is $4 \rm km\ s^{-1}Mpc^{-1}$, in the case of GW+AG+C fit. This is not good enough to prefer either the \textit{Planck} or the SHoES $H_0$, yet. More events are needed to reach their level of precision. However, in the future, we do not expect many events that have coincident detections of GW, afterglow light curve and centroid motion. Using the GW simulations from \citet{Petrov2022}, we generate the EM counterparts of more than a thousand binary neutron star events detected in O4 \citep{singerO42021} and O5 \citep{singerO52021}.  With the VLBI image resolution and sensitivity, we estimate that, both for O4 and O5, the rate of GW, afterglow light curve and centroid motion joint detections is a fraction of event per year.

To conclude, by introducing additional constraints based on astronomical observations, there is the potential to introduce systematic biases, that could affect the standard sirens measurements \citep{Chen2020, Nicolaou2020, GovreenSegal2023}. As we show in this work, the viewing angle in the EM modelling is affected by the type of data set used. For this reason, it is fundamental to include in the analysis all the messengers available, in order to have robust results.
At the moment, the uncertainty of the standard sirens methods is still too large with respect to the early or late-time Universe $H_0$s, but in the future attention should be taken, to avoid biases. For example, in a GW170817-like case, measurements at very late times could confirm or qualify as a systematic the milder decrease of the flux at late times. Regarding jet centroid studies, measures at both early and late times will be important to constrain its motion and its viewing angle. For these reasons, highly sensitive instruments are needed, like \textit{Athena} \citep{Piro2022} in the X-rays or SKA \citep[Square Kilometre Array,][]{Braun2019} in the radio band. In the distant future (mid 2030s), facilities like Next Generation VLA (ngVLA) will reach the mas resolution (or lower, \citealt{Beasley2019}), increasing the chances of detecting the motion of the relativistic jet.

\section*{Acknowledgements}

We thank the anonymous Referee for the very useful comments. Moreover, we thank Gabriele Bruni for the helpful discussion about radio telescopes present and future performances. We acknowledge support by the European Union horizon 2020 programme under the AHEAD2020 project (grant agreement number 871158). LP and GG also acknowledge support from MIUR, PRIN 2020 (grant 2020KB33TP) “Multimessenger astronomy in the Einstein Telescope Era" (METE). This work has been also supported by ASI (Italian Space Agency) through the Contract no. 2019-27-HH.0. Research at Perimeter Institute is supported in part by the Government of Canada through the Department of Innovation, Science and Economic Development and by the Province of Ontario through the Ministry of Colleges and Universities.
This research has made use of data or software obtained from the Gravitational Wave Open Science Center (gw-openscience.org), a service of LIGO Laboratory, the LIGO Scientific Collaboration, the Virgo Collaboration, and KAGRA. LIGO Laboratory and Advanced LIGO are funded by the United States National Science Foundation (NSF) as well as the Science and Technology Facilities Council (STFC) of the United Kingdom, the Max-Planck-Society (MPS), and the State of Niedersachsen/Germany for support of the construction of Advanced LIGO and construction and operation of the GEO600 detector. Additional support for Advanced LIGO was provided by the Australian Research Council. Virgo is funded, through the European Gravitational Observatory (EGO), by the French Centre National de Recherche Scientifique (CNRS), the Italian Istituto Nazionale di Fisica Nucleare (INFN) and the Dutch Nikhef, with contributions by institutions from Belgium, Germany, Greece, Hungary, Ireland, Japan, Monaco, Poland, Portugal, Spain. The construction and operation of KAGRA are funded by Ministry of Education, Culture, Sports, Science and Technology (MEXT), and Japan Society for the Promotion of Science (JSPS), National Research Foundation (NRF) and Ministry of Science and ICT (MSIT) in Korea, Academia Sinica (AS) and the Ministry of Science and Technology (MoST) in Taiwan.

\section*{Data Availability}

The data underlying this article will be shared on reasonable request to the corresponding author.



\bibliographystyle{mnras}
\bibliography{references} 


\appendix
\section{The degeneracy $d_L - \theta_v$ for the jet centroid motion}
\label{Sec:Appendix}

\begin{figure}
    \includegraphics[width=0.45\textwidth]{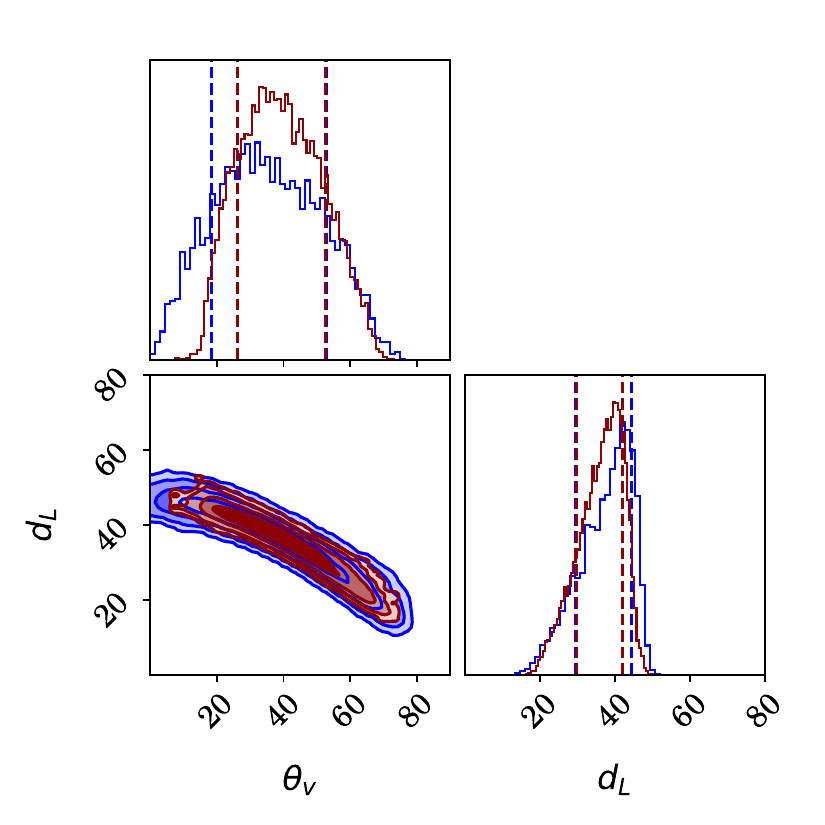}
    \caption{Contour plot of the viewing angle and luminosity distance. The contours represent the 68\%, 95\%, 99.7\% and 99.99\% probabilities. The blue contour lines represent the result from the GW fit, while the dark red lines represent the result from a fit using GW and centroid alone. }
     \label{Fig:dL_thetav_centr_gw}
\end{figure}

Unfortunately, also for the afterglow centroid motion there is a dependency on both the distance and the viewing angle. In fact, the magnitude of jet centroid motion in the sky can be estimated as 
\begin{equation}
    \Delta \theta_{\rm cent} = \Delta t_{\rm obs} \frac{d\theta_{\rm cent}}{dt_{\rm obs}}
\end{equation}
where $\Delta t_{\rm obs}$ represents the time period of the observations and $d\theta_{\rm cent} / dt_{\rm obs}$ the apparent velocity of the remnant in the sky. The latter can be written as \citep[see also][]{Mastrogiovanni2021}
\begin{equation}
    \frac{d\theta_{\rm cent}}{dt_{\rm obs}} = \frac{\beta_{\rm app} c}{d_A}
\end{equation}
where $d_A$ is the angular distance, $c$ is the speed of light and $\beta_{\rm app} = \beta \sin(\theta_v) / (1 - \beta \cos(\theta_v))$, where $\beta = \sqrt{1-1/\Gamma^2}$ and $\Gamma$ is the jet Lorentz factor. We can assume to observe the jet at the jet break, when the velocity is at its peak \citep{Granot2003, Ryan2023} and $\Gamma = 1/\theta_v$ (this is an overestimation, in reality, the velocity of the jet depends also on the energy of the jet and the circumburst medium, for example, a denser medium will produce a more luminous afterglow that moves slower, see \citealt{Ryan2023}).
For small $\theta_v$, $\beta\sim1$, so 
\begin{equation}
    \beta_{\rm app} \simeq \frac{2}{\theta_v}
\end{equation}
and the afterglow centroid motion in the sky can be written as 
\begin{equation}
    \Delta \theta_{\rm cent} \simeq \Delta t_{\rm obs} \frac{2 c}{\theta_v d_A}.
\end{equation} 
Also in this case, $d_L$ (and $d_A$) is a decreasing function of $\theta_v$. In the case of on-axis observers ($\theta_v<\theta_c$), the jet moves along the line of sight, so no displacement is visible. 
Because of the reasoning above, the centroid alone is not enough to break the GW degeneracy, indeed, a fit including only the centroid and the GW leads to no constraint on $\theta_v$ or $d_L$, see Fig.\,\ref{Fig:dL_thetav_centr_gw}.

\bsp	
\label{lastpage}
\end{document}